\documentclass[twocolumn]{emulateapj}
\usepackage{graphicx}
\usepackage{amssymb}
\usepackage{amsmath}
\usepackage{amssymb}

\def\aap {{A\&A}}

\def\apjl {{ApJL}}

\def\apjs {{ApJS}}

\newcommand{\FLASH}{{\sc flash} }
\newcommand{\ORION}{{\sc orion} }

\newcommand{\PLUTO}{{\sc pluto} }

\newcommand{\HII}{H{\sc ii} }
\newcommand{\Msun}{\,\rm{M}_\odot}

\renewcommand{\vec}[1]{\boldsymbol{#1}}
\newcommand{\grad}{\nabla}
\renewcommand{\div}{\grad \vec{\cdot} }

\shorttitle{Simulating the Formation of Massive Protostars}
\shortauthors{Klassen et al.}
\begin{document}
\bibliographystyle{apj}

\title{Simulating the Formation of Massive Protostars: I. Radiative Feedback and Accretion Disks}

\author{Mikhail Klassen}
\affil{Department of Physics and Astronomy, McMaster University \\ 1280 Main St.~W, Hamilton, ON L8S 4M1, Canada}
\email{klassm@mcmaster.ca}

\author{Ralph E.~Pudritz\altaffilmark{1}$^{\textrm{,}}$\altaffilmark{2}$^{\textrm{,}}$\altaffilmark{3}}
\affil{Department of Physics and Astronomy, McMaster University \\ 1280 Main St.~W, Hamilton, ON L8S 4M1, Canada}

\author{Rolf Kuiper\altaffilmark{3}}
\affil{Institute of Astronomy and Astrophysics, University of T\"{u}bingen \\ Auf der Morgenstelle 10, D-72076 T\"{u}bingen, Germany}

\author{Thomas Peters}
\affil{Max-Planck-Institut f\"{u}r Astrophysik \\ Karl-Schwarzschild-Str. 1, D-85748 Garching, Germany}

\and

\author{Robi Banerjee}
\affil{Hamburger Sternwarte, Universit\"{a}t Hamburg \\ Gojenbergsweg 112, D-21029 Hamburg, Germany}

\altaffiltext{1}{Origins Institute, McMaster University, 1280 Main St.~W, Hamilton, ON L8S 4M1, Canada}
\altaffiltext{2}{Institut f\"{u}r Theoretische Astrophysik, Albert-Ueberle-Str.~2, 69120 Heidelberg, Germany}
\altaffiltext{3}{Max Planck Institut f\"{u}r Astronomie, K\"{o}nigstuhl 17, 69117 Heidelberg, Germany}

\begin{abstract}
We present radiation hydrodynamic simulations of collapsing protostellar cores with initial masses of 30, 100, and 200$\Msun$. We follow their gravitational collapse and the formation of a massive protostar and protostellar accretion disk. We employ a new hybrid radiative feedback method blending raytracing techniques with flux-limited diffusion for a more accurate treatment of the temperature and radiative force. In each case, the disk that forms becomes Toomre-unstable and develops spiral arms. This occurs between 0.35 and 0.55 freefall times and is accompanied by an increase in the accretion rate by a factor of 2--10. Although the disk becomes unstable, no other stars are formed. In the case of our 100 and 200$\Msun$ simulation, the star becomes highly super-Eddington and begins to drive bipolar outflow cavities that expand outwards. These radiatively-driven bubbles appear stable, and appear to be channeling gas back onto the protostellar accretion disk. Accretion proceeds strongly through the disk. After 81.4 kyr of evolution, our $30 \Msun$ simulation shows a star with a mass of $5.48 \Msun$ and a disk of mass $3.3 \Msun$, while our $100 \Msun$ simulation forms a $28.8 \Msun$ mass star with a $15.8 \Msun$ disk over the course of 41.6 kyr, and our $200 \Msun$ simulation forms a $43.7 \Msun$ star with an $18 \Msun$ disk in 21.9 kyr. In the absence of magnetic fields or other forms of feedback, the masses of the stars in our simulation do not appear limited by their own luminosities.
\end{abstract}


\section{Introduction} \label{sec:introduction}

The formation of high-mass stars ($M \gtrsim 8$ $\Msun$) has been a subject of intense study and considerable controversy for several decades. Despite being slight in abundance, with only 2 or 3 high-mass stars for every 100, they exert enormous influence on the galaxies and host clusters in which they reside \citep{ZinneckerYorke2007}. Their extreme nature demands attention because the mechanisms of their formation are not immediately obvious. A defining feature of massive stars is that the timescale for their gravitational contraction, the Kelvin-Helmholtz time, is shorter than the timescale for accretion. This implies that the star begins nuclear burning before it has finished accreting to its final mass. During this phase, powerful feedback mechanisms due to radiation pressure and photoionization \citep{LarsonStarrfield1971} become important---mechanisms that are either absent or insignificant for low-mass star formation.

Observations exist for stars in excess of 150$\Msun$ \citep{Crowther+2010}. The 30 Doradus region inside the Large Magellanic Cloud contains at least 10 stars with initial masses in excess of 100 $\Msun$ \citep{Doran+2013}. There is no consensus on the precise upper mass limit for stars, although it may lie somewhere within 150--300 $\Msun$ \citep{Figer2005,Crowther+2010}. Assuming that such masses can be reached by accretion, then we must ask whether the accretion proceeds through a circumstellar disk alone, or additionally through other mechanisms. In addition, a large body of theoretical and numerical work shows that it is essential to adopt a three-dimensional treatment that includes a careful implementation of the radiation field. Many different theoretical and observational approaches have been undertaken to address this.

Massive stars are expected to reach the main sequence while still deeply embedded in the molecular envelopes \citep{Stahler+2000}. This makes observing high-mass protostars and their circumstellar disks challenging, although a growing number of candidates have been detected in recent years. \citet{BeltrandeWit2015} summarize the observations of accretion disks around luminous young stellar objects (YSOs). Where observations of circumstellar disks are available, their rotation rates appear consistent with Keplerian velocities. For massive protostars ($M_*$ = 8--30 $\Msun$), accretion disks generally range in size from a few hundred AU to a few thousand AU. These sizes are consistent with the disks formed in simulations presented in this paper.

Observations with the Atacama Large Millimeter/submillimeter Array (ALMA) are also revealing the properties of disks around massive protostars. Very recent observations highlighted the discovery of a Keplerian-like disk around a forming O-type star, AFGL 4176 \citep{Johnston+2015}. Model fits to line and continuum emission are consistent with a 25 $\Msun$ star surrounded by 12 $\Msun$ disk with a radius of 2000 AU. The velocity structure, as traced by CH$_3$CN, is consistent with Keplerian rotation. These observations are in excellent agreement with the results of our simulations of a 100 $\Msun$ protostellar core.

Early 1D calculations sought to estimate the upper mass limit for stars, given that the accretion flow onto a star should be halted at some critical luminosity. The Eddington Limit describes this maximal luminosity and depends on the ratio of radiative to gravitational forces, including the specific opacity of the gas and dust. Since the luminosity scales with the mass of the star more strongly than gravity, there should exist a natural upper mass limit for stars in the spherically-averaged case. Studies such as \citet{LarsonStarrfield1971, Kahn1974, YorkeKruegel1977} found upper mass limits around 20--40 $\Msun$, although such calculations are always sensitive to the choice of dust model. The interstellar medium is about 1\% dust by mass. Stellar radiation impinges on dust particles, imparting momentum. This momentum is transferred to the gas via drag forces. In 1D, a sufficiently luminous star (roughly 20 $\Msun$ or more) should be able to arrest and even reverse the accretion flow via radiation pressure alone, thus setting an upper mass limit for the star. How one chooses to model the dust opacity will affect the upper mass limit that results from these kinds of 1D calculations.

These theoretical 1D mass limits have been confirmed by simulations \citep{YorkeKruegel1977,Kuiper+2010b}. \citet{Kahn1974} suggested that the envelope of material around the protostar would likely fragment, leading to anisotropic accretion. The importance of geometry was later shown \citep{Nakano1989} and circumstellar disks highlighted as an accretion channel. These disks form easily in the presence of rotation. \citet{YorkeSonnhalter2002} studied slowly-rotating nonmagnetic cores in 2D simulations that assumed axial symmetry. Their simulations modeled the collapse of high-mass cores and included frequency-dependent radiation feedback handled via flux-limited diffusion (FLD). The effect of the disk is to channel radiation into the polar directions, a phenomenon known as the ``flashlight effect'' \citep{YorkeBodenheimer1999}. This radiation is also most responsible for radiative accelerations. Even so, \citet{YorkeSonnhalter2002} still only formed stars with masses $\lesssim 43\Msun$.

Radiatively-driven outflows are expected to form in the vicinity of massive stars. \citet{Krumholz+2009}, using a frequency-averaged flux-limited diffusion approach to radiative transfer, proposed that as the stars become more luminous, rather than blowing away all the gas enveloping the cavity, radiation would pierce through the cavity wall. Meanwhile, dense fingers of material would rain back down onto the massive protostar. This ``radiative Rayleigh-Taylor'' instability is analogous to the classical Rayleigh-Taylor instability, but with radiation taking the place of the lighter fluid. These instabilities may allow the star to continue accreting material. \citet{JacquetKrumholz2011} showed that these instabilities can arise under the right circumstances even around \HII regions.

This mechanism, however, may not be necessary to explain how massive stars achieve their prodigious masses. \citet{Kuiper+2010b} demonstrated that with a much more accurate treatment of the radiation field, including both FLD as well as ray-tracing of energy directly from the star, disk accretion alone was sufficient to achieve stellar masses larger than any previously achieved in a simulation.

The question becomes whether a classically two-dimensional fluid instability holds in a 3D environment, and whether it is appropriate to model the stellar radiation field as a fluid. \citet{Kuiper+2010a} decomposed the radiation field into a direct component and a diffuse component. The diffuse component, which consisted entirely of thermal radiation re-emitted by dust, was handled by an FLD solver. Meanwhile, the direct component consisted of the stellar radiation and was treated by a multifrequency raytracer. In simulations using this ``hybrid'' technique for radiation transfer, the radiative Rayleigh-Taylor instability was absent \citep{Kuiper+2012}. \citet{Kuiper+2010b} found that it was necessary to model and to resolve the dust sublimation front around the accreting protostar. With dust sublimation, matter becomes transparent to radiation in the extreme vicinity of the protostar. Rather than being blown away, it can now be accreted by the protostar.

These and subsequent simulations have shown that stars in excess of $100\Msun$ may form by disk accretion without the need for new fluid instabilities \citep{Kuiper+2010b,Kuiper+2011,Kuiper+2012}. However, these simulation also have their limitations. They were performed on grids with a spherical geometry and a sink cell fixed at the center. The advantage of this geometry is that it allowed for extremely high resolution near to the source. The radial grid resolution increases logarithmically towards the center. \citet{Kuiper+2010b} was able to resolve down to 1 AU in the vicinity of the star. Raytracing is also easy to implement in this geometry as rays travel only along the radial dimension. The disadvantage of this setup is that it only handles a single immovable source, fixed at the center. Fragmentation of the pre-stellar core was neglected, and the source cannot drift even as the disk becomes gravitationally unstable. While the resolution at small radii is very good, large-scale structure (particularly at large radii) are poorly resolved. These would be better handled in an adaptive mesh framework. 

The combined limitations of this body of work motivated our implementation of hybrid radiation transport on an adaptive, 3D Cartesian grid in \FLASH \citep{Klassen+2014}, first for the study of massive stars and their outflow cavities, then more general problems of star formation in clusters. In \citet{Klassen+2014} we demonstrated the accuracy of our method in solving the radiative transfer problem with a battery of benchmark tests, most of which were static. In this paper, we apply our method to the dynamical problem of following the gravitational collapse of a molecular cloud core to form a massive protostar. We also describe improvements and some necessary modifications that were made to the code in order for us to simulate massive star formation.

We performed three different computationally-intense simulations of protostellar cores collapsing gravitationally. These cores were of three different initial masses, 30 $\Msun$, 100 $\Msun$, and 200 $\Msun$, respectively. Their initial conditions emulated earlier studies of massive star formation, but that had different radiatively feedback techniques or different grid geometries. In our case, we formed only a single massive star in each simulation, with final masses of $5.48 \Msun$, $28.8 \Msun$, and $43.7 \Msun$. The simulations were run for 81.4 kyr (0.85 $t_{\textrm{ff}}$), 41.6 kyr (0.79 $t_{\textrm{ff}}$), and 21.9 kyr (0.59 $t_{\textrm{ff}}$), respectively. The formation of a massive protostar was accompanied by the formation of an accretion disk that grew in mass until becoming unstable, triggering a factor of 2--10 increase in the accretion rate, but which did not undergo further fragmentation into mutlpile stars. The strong radiative feedback from the massive protostars did not halt accretion, which continued through the disk, but launched radiatively-driven bubbles in the $100 \Msun$ and $200 \Msun$ simulations, in which the stars had super-Eddington luminosities.

Below we describe the setup of these simulations and their results in detail. Section \ref{sec:model} provides the background to the physical and numerical models we employ. Section \ref{sec:simulations} details the simulation setup. The results of these simulations is discussed in section \ref{sec:results}, while some nuances and caveats, as well as the direction of future simulations is discussed in section \ref{sec:discussion}. We summarize our results in section \ref{sec:conclusion}.

\section{Physical and Numerical Model} \label{sec:model}

\subsection{Radiation Transfer}

The details of our radiative transfer method are covered in \citet{Klassen+2014}, but we summarize the basics of the method here. The code was further developed to handle dynamical star formation calculations. These improvements are described in sections \ref{sec:developments} and \ref{sec:temp_iteration}. The radiation field is decomposed into a direct, stellar component and a diffuse, thermal component \citep{WolfireCassinelli1986,Murray+1994,EdgarClarke2003,Kuiper+2010a}, for which we can apply different methods that are better suited to each type of radiation:
\begin{equation}
\vec{F} = \vec{F}_* + \vec{F}_{r}
\end{equation}

The direct radiation flux from a protostar $F_*(r)$ measured a distance $r$ from the star is given by
\begin{equation}\label{eqn:stellar_flux}
F_*(r) = F_*(R_*) \left(\frac{R_*}{r}\right)^2 \exp\left(-\tau(r)\right),
\end{equation}
where $F_*(R_*) = \sigma T_{\mathrm{eff}}^4$ is the flux at the stellar surface $R_*$, where $\sigma$ is the Stefan-Boltzmann constant and $T_{\mathrm{eff}}$ the effective surface temperature of the protostar. The stellar flux is attenuated both geometrically with distance from the source, and by intervening material that is scattering or absorbing starlight. The latter effect is captured by the exponential term in Eq. \ref{eqn:stellar_flux}, where
\begin{equation}\label{eqn:optical_depth}
\tau(r) = \int_{R_*}^r \kappa(T_*,r^\prime) \rho(r^\prime) dr^\prime
\end{equation}
is the optical depth.

Raytracing is a method that is well-suited for treating the direct radiation field, determining for each grid cell the flux of photons arriving directly from the photosphere of any stars within the simulation. A raytracer making use of the fast voxel traversal algorithm \citep{AmanatidesWoo1987} is used to calculate optical depths to every cell in the computational domain, from which we calculate the local stellar flux. This flux depends on the integrated optical depth along the ray. The amount of stellar radiation absorbed by a grid cell depends on the local opacity,
\begin{equation}\label{eqn:irradiation}
\grad \cdot F_*(r) \approx - \frac{1-e^{-\kappa_P(T_*)\rho\Delta r}}{\Delta r} F_*(r)
\end{equation}
where $\kappa$ is the specific opacity and $\rho$ is the matter density. 

This absorbed energy enters the coupled energy equations as a source term:
\begin{eqnarray}
\label{eqn:eint_evol}
 \frac{\partial (\rho \epsilon)}{\partial t} &=& - \kappa_P \rho c \left(a_R T^4 - E_r \right) - \div \vec{F}_* \\
\label{eqn:erad_evol}
 \frac{\partial E_r}{\partial t} + \grad \cdot \vec{F}_r &=& + \kappa_P \rho c \left(a_R T^4 - E_r \right)
\end{eqnarray}

Here $E_r$ is the radiation energy, $\kappa_P$ is the Planck opacity, and $a_R$ is the radiation constant. The temperature in Equation \ref{eqn:eint_evol} is solved for implicitly (see Section \ref{sec:temp_iteration}), which is used in equation \ref{eqn:erad_evol}. Equation \ref{eqn:erad_evol}, in turn, is solved implicitly using the flux-limited diffusion approximation \citep{LevermorePomraning1981}, 
\begin{equation}\label{eqn:fld_approximation}
\vec{F}_r = - \left(\frac{\lambda c}{\kappa_R \rho}\right) \grad E_r, 
\end{equation}
where $\kappa_R$ is the Rosseland mean opacity, $c$ is the speed of light, and $\lambda$ is the flux-limiter. Different choices for the flux limiter have been made in the literature, and are based on slightly different assumptions about the angular distribution of the specific intensity of the radiation field. The flux limiter always takes on values between $0$ and $\frac{1}{3}$ \citep{LevermorePomraning1981,TurnerStone2001}.

We use the \citet{LevermorePomraning1981} flux limiter, one of the most commonly used,
\begin{equation}
\lambda = \frac{2+R}{6+3R+R^2},
\end{equation}
with
\begin{equation}
R = \frac{|\grad E_r|}{\kappa_R \rho E_r}.
\end{equation}
Another popular choice of flux limiter is the one by \citet{Minerbo1978}.

The advantage of FLD methods is their speed and relative accuracy in regions of high optical depth. And while massive stars form in very dense envelopes of molecular gas \citep{Sridharan+2002,Hill+2005,Klein+2005,Beltran+2006}, within the dust sublimation front, or inside the radiatively-driven outflow cavities that form around massive stars, the radiation field is highly anisotropic. The FLD approximation, however, assumes that flux is in the direction antiparallel to the gradient of the radiation energy $E_r$. FLD methods can become less accurate in regions transitioning from optically thin to optically thick, or in regions with highly anisotropic radiation fields.

The addition of stellar sources of radiation in FLD-only radiation transfer codes is often handled via an isotropic source term that considers the luminosity of the star:
\begin{equation}\label{eqn:fld_w_sinks}
\frac{\partial E_r}{\partial t} + \div \left(\frac{\lambda c}{\kappa_R \rho} \grad E_r\right) = \kappa_P \rho c \left(a T^4 - E_r \right) + \sum_n L_n W(\mathbf{x} - \mathbf{x}_n)
\end{equation}
The coupling of radiation energy from the stars to the gas is mediated by a weighting function centered on the stellar locations and appears as a source term in the radiation energy diffusion equation. The weighting function $W$, which might be a spherical Gaussian function, mediates the radiation energy contributed by a sink particle $n$, located at $\mathbf{x}_n$, which has a luminosity $L_n$. The luminosity may be computed via a model for protostellar evolution or an estimate based on a zero-age main sequence (ZAMS) star, but if the weighting function is symmetric, then the radiation is still administered isotropically. Unfortunately, these regions, if circumstellar disks are present or assumed, are often the grid cells where the radiation field can be very anisotropic.

Using solely a raytracer has major disadvantages as well. For computational efficiency, scattering and non-stellar source terms are often neglected, which implies that this method is accurate only in optically thin regions where the radiation field is dominated by stellar sources. In reality, the presence of optically thick regions means radiation is absorbed, re-emitted at lower energy, or scattered by particles. This occurs especially in the envelopes around massive protostars, where the gas and radiation become tightly coupled. For reasons of technical complexity or computational cost, raytracers have typically been avoided for dynamical star formation calculations, though Monte Carlo raytracers are sometimes used in post-hoc analysis of data. In numerical simulations, diffusion codes are popular, which perform well under conditions of tight matter-radiation coupling. That said, \citet{Buntemeyer+2015} re-implemented a characteristics-based raytracer in \FLASH to calculate the mean radiation intensity at each cell using the Accelerated Lambda Iteration (ALI) approach \citep[see][]{ALI1995}, allowing for dynamical star formation simulations without a major loss of accuracy. 

Monte Carlo raytracers are another species of raytracer worth mentioning because of their popularity and accuracy. These follow the path of many photon packets and solve the full radiative transfer problem including many generations of photon absorption, re-emission, and scattering. With enough simulated photons, highly accurate estimations of temperature may be made. However, these methods are so computationally expensive that they are not typically implemented in dynamical calculations. There are exceptions, however. Dynamical simulations with Monte Carlo radiative transfer were performed by \citet{Harries+2014,Harries2015}.

By decomposing the radiation field into a direct component and a diffuse component, and computing the first via a raytracer and the second via the diffusion approximation, the worst flaws of each method are minimized. It is not that different radiative transfer methods are active in different parts of the grid. Rather, at each location in the grid, the raytracer is computing the incident stellar flux, though it may be extremely attenuated far from any sources. Meanwhile, the FLD solver is also operating on the entire grid to diffuse the thermal radiation, with the incident stellar flux appearing as a source term. This is how the two methods are linked. As a result, we compute a more accurate equilibrium gas temperature and radiative force. The flashlight effect is enhanced and circumstellar disks more effectively shield incoming material from stellar radiation.

The validity of this hybrid radiative transfer approach and its application to various problems of star formation has been demonstrated by \citet{Kuiper+2010a,Kuiper+2012,KuiperKlessen2013,Klassen+2014} and the method has been implemented in other adaptive mesh codes \citep{RamseyDullemond2015}.

\subsection{Further developments of the hybrid radiation transfer code}\label{sec:developments}

Since the publication of \citet{Klassen+2014}, we have been applying our code to gravitational collapse calculations. One drawback of the previous implementation was that our hydrodynamics solver was not taking into account any dynamical effects resulting from gradients in the radiation field. The only direct consequence of the diffuse thermal radiation field was the addition of an isotropic radiation pressure, $P_{\textrm{rad}} = E_r / 3$, to the total pressure.

We have now switched our hydro solver from a split method to the unsplit hydro solver included in \textsc{flash}, with modifications for
handling radiation in the flux-limited diffusion approximation that follow the equations of \citet{Krumholz+2007b}, which are similar to the implementation described in \citet{Zhang+2011}. Our simulations were done with a pre-release version of this modified unsplit hydro solver. It has since been released as part of \FLASH version 4.3. Diffuse thermal radiation no longer contributes an isotropic radiation pressure to the total pressure. Instead, the gradient of the radiation energy density and the flux limiter are considered in the momentum equation (see Equation \ref{eqn:momentum_conservation} below).

Directionally split solvers sequentially consider rows of cells in 1D and solve the Riemann problem at cell boundaries. The $x$, $y$, and $z$ dimensions are solved in turn. Unsplit solvers consider a larger kernel of cells and can include more cross-terms in the solving of the hydrodynamic equations. Unsplit solvers are better at minimizing grid artifacts and preserving flow symmetries.

The relevant set of radiation hydrodynamic equations follows the mixed-frame formulation of \citet{Krumholz+2007b}, keeping terms up to $O(v/c)$ and dropping terms insignificant in the streaming, static diffusion, and dynamic diffusion limits. Mixed-frame implies that the radiation quantities are written in the lab frame, but fluid quantities, in particular the fluid opacities, are written in the comoving frame. The approach begins with writing the radiation hydrodynamics equations in the lab frame, applying the flux-limited diffusion approximation in the comoving frame, and then transforming back into the lab frame and retaining terms to order $v/c$. The mixed-frame approach is advantageous because it conserves total energy and is well-suited for AMR codes. For a detailed derivation, see \citet{Krumholz+2007b}.

In practice, it is not exclusively the domain of the hydro solver to operate on the set of equations below. Rather, like most similar astrophysics codes, \FLASH employs operator-splitting for handling diffusion and source terms. The most important feature of the solver modifications mentioned above consists in making the hydro solver ``flux-limiter aware'', rather than restricting the effect of flux limiting to the diffusion solver.

\FLASH solves the following equations for the mass, momentum, internal energy of the gas and radiation field energy density:
%
%
\begin{eqnarray}
\label{eqn:mass_conservation}
&\frac{\partial \rho}{\partial t} + \div \left(\rho \vec{v}\right) = 0 \\
\label{eqn:momentum_conservation}
&\frac{\partial \left(\rho \vec{v}\right)}{\partial t} + \div \left( \rho \vec{v}\vec{v}\right) + \grad p + \lambda \grad E_r = \frac{\kappa_P}{c} \vec{F}_* - \rho\grad\Phi \;\\
\label{eqn:energy_transport}\nonumber
&\frac{\partial \left(\rho e\right)}{\partial t} + \div \left( \rho e \vec{v} + P\vec{v}\right) = -\kappa_P \rho c \left(a_R T^4 - E_r \right) \\ &+ \lambda \left( 2 \frac{\kappa_P}{\kappa_R} - 1 \right) \vec{v} \cdot \grad E_r - \div \vec{F}_* \\
\label{eqn:radiation_energy_transport}\nonumber
&\frac{\partial E_r}{\partial t} - \div \left( \frac{c \lambda}{\kappa_R \rho} \grad E_r \right) = - \div \left( \frac{3-f}{2} E_r \vec{v} \right) \\ &+\kappa_P \rho c \left(a_R T^4 - E_r \right) - \lambda \left( 2 \frac{\kappa_P}{\kappa_R} - 1 \right) \vec{v} \cdot \grad E_r
\end{eqnarray}

In the above equations, $\rho$, $v$, $p$, $e$, and $T$ are the gas density, velocity, pressure, specific internal energy, and temperature, while $E_r$, $\lambda$, $c$, and $a_R$ are the radiation energy density, the flux limiter, the speed of light, and the radiation constant. $\Phi$ is the gravitational potential. $\kappa_P$ and $\kappa_R$ are the Planck and Rosseland mean opacities. $f$ denotes the Eddington factor,
\begin{equation}\label{eqn:eddington_factor}
f = \lambda + \lambda^2 R^2.
\end{equation}

In the optically-thick limit, the flux limiter $\lambda$ and the Eddington factor $f$ both approach 1/3, whereas in the optically-thin limit these approach 0 and 1, respectively.

This set of equations becomes equivalent to the one solved in \citet{Klassen+2014} under the following simplifications: first, replace all ocurrences of $\lambda$ and $f$, except in the diffusive term of (\ref{eqn:radiation_energy_transport}), by the value appropriate for the diffusion limit, 1/3; second, omit the Lorentz transformation terms.

The Appendix of \citet{Klassen+2014} also no longer applies, since we have since switched to a two-temperature model. \FLASH's original flux-limited diffusion solver was designed to handle separate electron and ion matter species, and the Appendix to \citet{Klassen+2014} described how equilibrate their temperatures. We have done away with this distinction and deal only with a single matter species (molecular gas).

We tested the flux-limiter-aware hydro solver on a classic benchmark for the combined effects of radiation and fluid motion, the 1-D critical radiation shock test. In this setup, gas flows at a given speed to the left and strikes a wall, which we implement as a reflecting boundary condition. This creates a shock that travels upstream. The compressed gas is heated and radiates energy upstream. The radiation field is coupled to the gas, and so the incoming material is preheated before encountering the shock. Semi-analytic estimates by \citet{MihalasMihalas84} allow for the comparison of measured temperatures.

The test is described in greater detail in \citet{Klassen+2014}. We use a ratio of specific heats $\gamma = 5/3$ and a mean molecular weight $\mu = 0.5$. In \citet{Klassen+2014}, we obtained adequate results with our radiation code on this benchmark, but our preshock temperature for the subcritical case where overshooting the semi-analytic value considerably (by $\sim20\%$).

\begin{figure}
\includegraphics[width=88mm]{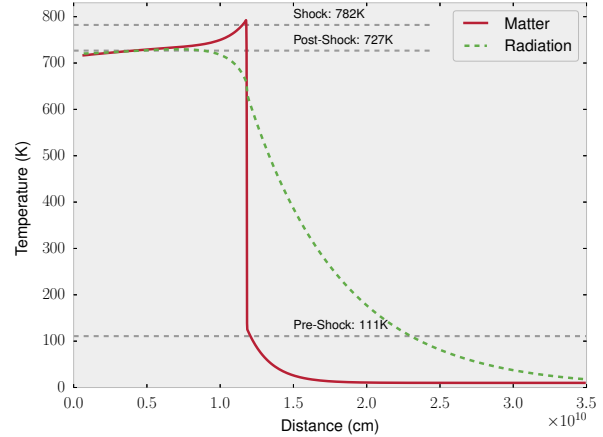}
\caption{Temperature profile of a 1D subcritical radiating shock with a gas velocity of $v = 6$ km/s at a time $t = 5.8 \times 10^4$ s. The red solid line and green dashed line indicate the matter and radiation temperatures, respectively. Gray horizontal dashed lines were added to indicate the semi-analytic estimates for the temperature in the preshock, shock, and postshock regions. Stated temperatures are the semi-analytic estimates for the various regions surrounding a subcritical shock.\\}
\label{fig:subcritical_shock}
\end{figure}

Figure \ref{fig:subcritical_shock} gives the revised calculation using the current version of hydrodynamics and radiation modules. The postshock temperature of $T_2 = 717$ K agrees to within 2\% of the semi-analytic estimate. The temperature ($T_+ = 792$ K) inside the shock agrees with the analytic value of $782$ K to within 2\%. Finally, the preshock temperature, where our code previously had the most difficulty matching semi-analytic estimates, is $T_- = 125$ K, compared to an expected value of $111$ K---a difference of 13\%.

In simulations of supercritical shocks, the revised code showed no major improvements over the results of \citet{Klassen+2014}.

\subsection{Temperature Iteration}\label{sec:temp_iteration}

In \citet{Klassen+2014}, equation 21, we expressed the implicit temperature update as
\begin{small}
\begin{equation}\label{eqn:temp_update}
T^{n+1} = \frac{3 a_R \alpha \left(T^n\right)^4 + \rho c_V T^n + \alpha E_r^{n+1} - \Delta t \div \vec{F}_*}{\rho c_V + 4 a_R \alpha \left(T^n\right)^3},
\end{equation}
\end{small}
where we have used $\alpha \equiv \kappa_P^n \rho c \Delta t$. The approach follows \citet{Commercon+2011} and involves a discretization of the coupled energy evolution equations, followed by a linearization of the temperature term, $(T^{n+1})^4$. We have now added a Newton-Raphson iteration to the temperature update (Eq. \ref{eqn:temp_update}), which was not previously necessary because all of the benchmark tests in \citet{Klassen+2014} involved either very short timesteps or were static irradiation tests. In a star formation simulation, the time steps are much longer, and the temperature in a cell could change by a significant amount (more than 10\%) in a single time step. The emission term scales as $T^4$ and the opacities are also temperature-dependent. When the equations of internal and radiation energy transport were discretized, we retained the opacity $\kappa$ from the previous time index $n$ because $\kappa = \kappa(T^{(n)})$ is a function of temperature $T$ and the updated temperature $T^{(n+1)}$ had yet to be determined. Provided the temperature does not change dramatically, this approach is acceptable, but in our later dynamical calculations involving star formation, longer timesteps and strong irradiation from the massive protostar meant that Equation (\ref{eqn:temp_update}) needed to be applied via a Newton-Raphson iteration. We thus modified the code to cycle the temperature update until the changes fell within safe tolerances: a minimum of 3 iterations, a minimum relative error of $10^{-5}$, and a minimum absolute error of $1$ K. 

\subsection{FLASH}

We have implemented the above radiative transfer method in the publicly available general-purpose magnetohydrodynamics code \FLASH \citep{Fryxell+2000, Dubey+2009}, now in its fourth major version. \FLASH solves the equations of (magneto)hydrodynamics using the high-order piecewise-parabolic method of \citet{ColellaWoodward1984} on an Eulerian grid with adaptive mesh refinement (AMR) by way of the PARAMESH library \citep{MacNeice+2000}. Flux-limited diffusion is handled via a general implicit diffusion solver using a generalized minimum residual (GMRES) method \citep{GMRES} that is part of the HYPRE libraries \citep{HYPRE} for parallel computation.

\subsection{Pre-main sequence evolution and sink particles}

\FLASH has a modular code framework and can thus be easily extended with new solvers and physics units. It has been modified to handle Lagrangian sink particles for representing stars \citep{Federrath+2010} or star clusters \citep{Howard+2014}. These particles exist within code grid and interact with the gas in the grid via mutual gravitation. A sink particle represents a region in space that is undergoing gravitational collapse. They were first implemented in grid codes because it was computationally infeasible to continue resolving the gravitational collapse of gas down to stellar densities. Instead, regions of runaway collapse are identified and sink particles created in their place. We understand that at the center of a sink particle resides a protostar that accretes from the gas reservoir around it---any gas above a threshold density is added to the mass of the particle (see details in section \ref{sec:sink_particles}).

The pre-main sequence evolution of a protostar involves changes in stellar structure that result in changes in radius, effective surface temperature, and luminosity. For massive stars, they continue to grow and accrete material even after they have reached the main sequence. When modeling the radiative feedback of stars, it is important to capture all of these transitions. The amount of radiation injected into the simulation depends heavily on how the stellar properties are modeled.

\citet{Klassen+2012a} explored the pre-main-sequence evolution of protostars and a protostellar model was implemented in \FLASH based on the one-zone model described in \citet{Offner+2009}. The star is modeled as a polytrope, and the stellar properties (mass, radius, luminosity, effective surface temperature) are evolved self-consistently as the star accretes material. The model was designed with high accretion in mind and is robust even when accretion rates are highly variable or episodic. We demonstrated in \citet{Klassen+2012a} that one-zone protostellar evolution models can handle high and extremely variable accretion rates. We compared our results to \citet{HosokawaOmukai2009}, who had a more sophisticated treatment of the stellar structure evolution for massive protostars with high accretion rates and showed acceptable agreement (but see \citealp*{HaemmerlePeters2016} for a critical assessment of high-mass pre-main-sequence evolution models). \citet{Krumholz+2009} also use protostellar evolution model. While not perfect, these models represent a more accurate treatment of the protostellar evolution, especially the stellar radius and surface temperature, than the traditional approach of fitting to a ZAMS model. These stellar properties are of great importance when calculating radiative feedback. The evolution of the sink particles formed in our simulations is described in section \ref{sec:star_formation_results}.

In each simulation, we witnessed the sink particle form at the center of the simulation volume. Given the comparable disk and stellar masses during the early evolution of the system, when the disk did become gravitationally unstable, the sink particle was perturbed from its location at the center of the simulation volume. The total displacement over the course of the simulation was less than 1000 AU in each simulation, almost all of which was in the plane of the disk. The higher the mass of the initial protostellar core, the less the sink particle moved.

\subsection{Opacity and Dust Model}

For radiative transfer calculations, we must consider the properties of interstellar dust. In the typical interstellar medium, the dust-to-gas ratio is about 1\% by mass. We use the opacity tables of \citet{DraineLee1984} that compute the optical properties of interstellar dust composed of graphite and silicates. These opacities we average over frequency to compute temperature-dependent dust opacities. In Figure 2 of \citet{Klassen+2014} we showed the matter temperature-dependent Rosseland and Planck mean dust opacities. The former is appropriate for treating radiation in the diffusion approximation, whereas we use the latter when calculating the absorption or emission of radiation by dust. We use the same opacities in this paper.

We assume $T_{\textrm{gas}} = T_{\textrm{dust}} = T_{\textrm{matter}}$, which is appropriate for most simulations of massive star formation, and which most FLD simulations assume. We distinguish between matter and radiation ($T_r = (E_r/a_R)^{0.25}$) temperatures, which exchange energy via emission and absorption. A third temperature that we calculate is the surface effective temperature of the star, which is used to calculate the direct radiative flux from the star.

The opacities are calculated for temperatures ranging between 0.1 K to 20000 K. However, in regions of hot gas, especially surrounding the protostar, it is necessary to consider dust sublimation, which modifies the opacity. We account for this in the form of a correction factor to the opacity, $\kappa(\vec{x}) = \epsilon \left(\rho,T_\mathrm{gas}\right) \kappa(T)$. To estimate the temperature at which dust begins sublimating, we use the formula by \citet{IsellaNatta2005},

\begin{equation}\label{eqn:dust_evap_temperature}
T_{\mathrm{evap}} = g \rho^\beta,
\end{equation}
where $g = 2000$K, $\beta = 0.0195$, and $\rho$ is the gas density. This formula is based on a power-law approximation to the dust properties as determined by \citet{Pollack+1994}. The correction factor we use is a smoothly-varying function of density and gas temperature, as in \citet{Kuiper+2010b}:

\begin{equation}\label{eqn:dust_evap_correction}
\epsilon(\rho,T) = 0.5 - \frac{1}{\pi}\left(\frac{T(\vec{x}) - T_{\mathrm{evap}}(\vec{x})}{100}\right)
\end{equation}

\begin{figure}
\includegraphics[width=88mm]{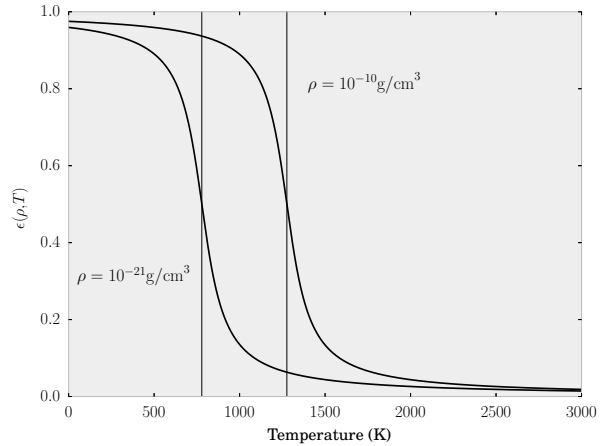}
\caption{Dust opacity correction to account for dust sublimation in regions of high gas temperature. Here we show the correction as a function of gas temperature for two different gas densities, $\rho = 10^{-21}$ and $10^{-10}$ g/cm$^3$. The vertical lines indicate the sublimation temperatures $T_{\textrm{evap}}$ for the given gas densities. The approach is identical to \citet{Kuiper+2010b}.}
\label{fig:dust_evaporation}
\end{figure}

Figure \ref{fig:dust_evaporation} shows the form of this function for two different gas densities, one high and one low. The vertical lines in the figure show the sublimation temperature $T_{\mathrm{evap}}$ for the two different densities, computed via equation \ref{eqn:dust_evap_temperature} and used in equation \ref{eqn:dust_evap_correction} to scale the dust opacity. Dust sublimation plays a crucial role in the inner regions of the disk nearest to the star. It both allows material to continue accreting and radiation an escape channel into the lower-density polar regions around the protostar.

One minor artifact of this method is that if gas temperatures drop, dust spontaneously reforms. More sophisticated dust treatments, including dust formation times or gas mixing, are theoretically possible, but would not substantively change the results of our simulation. The only regions affected by dust sublimation are the grid cells immediately surrounding the star, where the gas is sufficiently hot.

\subsection{Radiation Pressure}

The radiation field imparts momentum on the gas via dust-coupling. Dust grains absorb radiation and receive an impulse. Drag forces then transfer momentum to the gas, and we assume that the coupling is strong enough that we do not need to consider the dust and gas as separate fluids, although there are some circumstances, such as in champagne flows of \HII regions \citep{Tenorio-Tagle1979}, where the dust and gas can begin decoupling \citep[see, e.g.,][]{Ochsendorf+2014}. 

Because our radiation field is decomposed into direct and diffuse components, so also does our radiation pressure have direct and diffuse components. The implementation in \FLASH and tests of accuracy are detailed in \citep{Klassen+2014}. To summarize, the body force exerted by the direct radiation field is given by equation 31 from \citet{Klassen+2014},
\begin{equation}\label{eqn:radiation_force_density}
f_{\mathrm{rad}} = \rho \kappa_P(T_*) \frac{F_*}{c} = - \frac{\div \vec{F}_*}{c},
\end{equation}
where $T_*$ is the temperature of the photosphere, which is the characteristic temperature of the direct radiation field. We see that the force is proportional to the absorbed radiation energy in a grid cell. The stellar flux, also, depends on the optical depth (see Equation \ref{eqn:stellar_flux}). In the gray radiation approximation, the optical depth is integrated using the stellar surface temperature for the opacity calculation (see Equation \ref{eqn:optical_depth}). This is because interstellar dust has a strong frequency dependence and most of the energy and momentum is carried by the high-frequency radiation. The result is that most of the energy and momentum from the direct radiation is deposited in an area close to the star. However, as the dust sublimates and an optically thin cavity begins to form, more of the direct radiation is able to escape into the polar direction.

For the diffuse (thermal) radiation field, the updated hydrodynamics solver incorporates the effects of radiation on the momentum of the gas.

While we currently use a frequency-averaged opacity model, we do not expect there to be large deviations in radiation pressure when compared to a multifrequency approach. The reason for this is that while low frequency radiation can penetrate further into the gas, it also carries much less energy compared to the higher-frequency components.

\section{Numerical Simulations} \label{sec:simulations}

Our goal was to emulate the initial conditions of the simulations of \citet{Krumholz+2009} and \citet{Kuiper+2010b,Kuiper+2011} to determine what effect our improved numerical and physical treatment has on the massive star formation problem. Each paper sought to address the question of how massive stars continued to accrete despite their enormous luminosities. As noted in the introduction, the main difference between their methods was the type of radiative transfer code used, with the former using an FLD-based scheme on a Cartesian AMR grid in \ORION and the latter using a hybrid raytrace/FLD scheme implemented in \PLUTO on a fixed, spherical grid. \citet{Krumholz+2009} showed simulations in which a radiative Rayleigh-Taylor instability fed material back onto the star and disk, whereas \citet{Kuiper+2012} argued that dust sublimation and a more accurate treatment of the radiation field in proximity to the star allows for a stable radiatively-driven cavity to form around the star, and for material to continue accreting through the disk. In the latter radiation scheme, since the stellar radiation is first carried by a raytracer, it is the optical depth along each ray that determines the star's impact. Pure FLD schemes use a kernel centered on the star to mediate the radiative energy injection, typically isotropically. In both cases, it was found that radiation pressure did not imply an upper mass limit for star formation.

As in the mentioned papers, the volume we simulate has a side length of 0.4 pc, with an central core of radius 0.1 pc residing inside. The core is initially cold ($T_0 = 20$K) and the molecular gas and dust has a matter density profile that follows a power-law,
\begin{equation}
\rho(r) \propto r^{-p},
\end{equation}
with $p = 1.5$. The density is scaled by a multiplicative constant to give the desired total core mass (30, 100, or 200 $\Msun$). To avoid a cusp, a quadratic smoothing function over the central $6$ cells smooths out the singularity. 

The core is initially in solid-body rotation. We choose an angular rotation rate to match that used in \citet{Kuiper+2010b}, $\Omega_0 = 5 \times10^{-13}$ s$^{-1}$. \citet{Krumholz+2009} set their rotation so that the ratio of rotational kinetic energy to gravitational binding energy,
\begin{eqnarray}
\nonumber
\beta_{\mathrm{rot}} &= \frac{\frac{1}{2} I \Omega^2}{U_{\mathrm{grav}}} = \frac{\left(\frac{3-p}{5-p}\right)M_{\mathrm{core}}R^2 \Omega^2 /3}{\left(\frac{3-p}{5-2p}\right) G M_{\mathrm{core}}^2 / R} \\ \label{eqn:beta_rot}
&= \left(\frac{5-2p}{5-p}\right)\frac{R^3\Omega^2}{3 G M_{\textrm{core}}} ,
\end{eqnarray}
is 2\% for a core with a power-law profile with index $p$. Our rotational energy ratios are higher (see Table \ref{table:simulation_parameters}). Higher rotation rates result in larger accretion disks, which should be more prone to fragmentation at larger radii. The rotation rates of disks eventually settle into Keplerian motion around the central star. 

\subsection{Sink particles}\label{sec:sink_particles}

Our simulations are initialized without any stars present, but sink particles are allowed to form naturally according to their formation criteria \citep{Federrath+2010} and act as the ray source location in the raytracing code. Sink particles interact with the gas via mutual gravitation and are not held fixed at the center of the simulation volume, a key difference between our simulation and those of \citet{Kuiper+2010b}, wherein the source was held fixed at the center of the spherical grid and represented by a sink cell. Material crossing the inner grid boundary at $r_{\mathrm{min}} = 10$ AU was considered accreted onto the star. In \cite{Krumholz+2009} the sink particles were allowed to form in regions that were Jeans-unstable and undergoing gravitational collapse (as in our case), according to the sink particle algorithm in \ORION \citep{Krumholz+2004}. 

In order to form a star, gas must be Jeans unstable. The \citet{Truelove+1997} criterion states that the Jeans length must be resolved by at least four grid cells, this imposes a natural density threshold for sink particle formation. If the sink radius, $r_{\textrm{sink}}$, is equal to the Jean's length, $\lambda_J = (\pi c_s^2 / (G\rho))^{1/2}$, then the threshold density becomes
\begin{equation}\label{eqn:density_threshold}
\rho_{\textrm{thresh}} = \frac{\pi c_s^2}{4 G r_{\textrm{sink}}^2},
\end{equation}
where the factor of $4$ in the denominator comes from the Truelove criterion and the sink radius is defined by the grid scale, $r_{\textrm{sink}} = 2.5 \Delta x$. Given our grid resolution of $\Delta x \approx 10$ AU, this results in a threshold density of $\rho \sim 10^{-14}$ g/cm$^3$ for all of the simulations described in this paper.

The sink particles in \citet{Krumholz+2009} followed an implementation described in \citet{Krumholz+2004}, with the threshold density and Jeans criterion determining sink particle creation. The sink particle algorithm of \citet{Federrath+2010} that we use is stricter, checking also for convergent flow, a gravitational potential minimum, and a negative total energy ($E_{\textrm{grav}} + E_{\textrm{th}} + E_{\textrm{kin}} < 0$) within a control volume.

\subsection{Resolution}

In order to resolve the dust sublimation region around the accreting star, we operate with 11 levels of refinement on our Cartesian grid, resulting in a grid resolution of $\Delta x_{\mathrm{cell}} = L / 2^{l_{\mathrm{max}}+2} = 10.07$ AU. \citet{Krumholz+2009} used a similar resolution in their simulations.

For our refinement criterion we choose to resolve the Jeans length with 8 grid cells, that is
\begin{equation}
\lambda_J = \sqrt{\frac{15 k_B T}{4 \pi G \mu \rho}} \ge 8 \Delta x_{\mathrm{cell}},
\end{equation}
where $k_B$ is Boltzmann's constant, $T$ is temperature, $G$ is Newton's constant, $\mu = 2.33$ is the mean molecular weight in our simulation, and $\rho$ is the gas density.

Additionally, any block containing the sink particle is refined to the highest level.

\subsection{Turbulence}

Star-forming environments inside molecular clouds are turbulent \citep{Larson1981, Evans1999}, and turbulence plays a crucial role in regulating star formation \citep{MacLowKlessen2004,McKeeOstriker2007,FederrathKlessen2012}. However, in this paper we are studying the role of radiation feedback in massive star formation, revisiting important simulations in the scientific literature, but without some of the previous limitations. It is essential, then, that other complexities, such as the inclusion of initial turbulence and magnetic fields, be left out until the `basic case' has been treated and fully studied. In forthcoming papers, we aim to study star formation in more realistic environments featuring turbulence and magnetic fields.

\subsection{Summary}

We performed a total of three supercomputer simulations, varying the initial core mass across these. We selected initial cores with masses of $30 \Msun$, $100 \Msun$, and $200 \Msun$. We kept all other parameters consistent between these simulations, including initial temperature, rotation rate, resolution, etc.

\begin{table*}[!htb]
\caption{Simulation parameters}\label{table:simulation_parameters}
\tablenum{1}
\centering
\begin{tabular*}{0.9\textwidth}{@{\extracolsep{\fill} } llcccc}
\hline\noalign{\smallskip}
\multicolumn{6}{c}{Physical simulation parameters} \\
\noalign{\smallskip}\hline\noalign{\smallskip}
Parameter & & & {\tt M30} & {\tt M100} & {\tt M200}  \\
\noalign{\smallskip}\hline\noalign{\smallskip}
Cloud radius        & [pc]        & $R_0$                  & 0.1 & 0.1 & 0.1 \\
Power law index &     & $p$   & 1.5 & 1.5 & 1.5 \\
Total cloud mass    & [$\Msun$]   & $M_{\textrm{tot}}$     & 30 & 100 & 200 \\
Number of Jeans masses   &  & $N_{\textrm{J}}$ & 13.03 & 79.33 & 224.37 \\
Peak (initial) mass density   & [g/cm$^3$]  & $\rho_c$ & $2.58 \times 10^{-15}$ & $8.84 \times 10^{-15}$ & $1.77 \times 10^{-14}$ \\
Peak (initial) number density & [cm$^{-3}$] & $n_c$    & $6.62 \times 10^{8}$ & $2.27 \times 10^{9}$   & $4.54 \times 10^{9}$   \\
Initial temperature   & [K]    & $T$   & 20 & 20   & 20 \\
Mean freefall time & [kyr] & $t_{\textrm{ff}}$ & 95.8 & 52.4 & 37.0 \\
Sound crossing time (core)& [Myr] & $t_{\textrm{sc}}$ & 0.569 & 0.569 & 0.569 \\
Rigid rotation angular frequency & [s$^{-1}$] & $\Omega_{\textrm{rot}}$ & $5.0 \times 10^{-13}$ & $5.0 \times 10^{-13}$ & $5.0 \times 10^{-13}$ \\
Rotational energy ratio & & $\beta_{\textrm{rot}}$ & 35.1 \% & 10.5 \% & 5.3 \% \\
\noalign{\smallskip}\hline\noalign{\smallskip}
\multicolumn{6}{c}{Numerical simulation parameters} \\
\noalign{\smallskip}\hline\noalign{\smallskip}
Simulation box size & [pc] & $L_{\textrm{box}}$ & 0.40 & 0.40 & 0.40  \\
Smallest cell size & [AU] & $\Delta x$ & 10.071 & 10.071 & 10.071 \\
\noalign{\smallskip}\hline\noalign{\smallskip}
\multicolumn{6}{c}{Simulation outcomes} \\
\noalign{\smallskip}\hline\noalign{\smallskip}
Final simulation time & [kyr] & $t_{\textrm{final}}$ & 81.4 & 41.6 & 21.9 \\
Number of sink particles formed& & $n_{\textrm{sinks}}$ & 1 & 1 & 1 \\
Final sink mass & [$\Msun$] & & 5.48 & 28.84 & 43.65  \\
\hline \\
\end{tabular*}
\end{table*}

The initial conditions and final sink particle properties are summarized in Table \ref{table:simulation_parameters}. In addition to the simulations listed, we performed supplementary simulations of the 30 $\Msun$ and 100 $\Msun$ protostellar core setups at lower rotation rates so that $\beta_{\textrm{rot}} = 2\%$ (see Equation \ref{eqn:beta_rot}). These lower rotation results did not change our conclusions.

\section{Results}\label{sec:results}

In this section we discuss the results from the three simulations described in the section above and summarized in Table \ref{table:simulation_parameters}.

\subsection{General description: different phases of massive star formation}\label{sec:star_formation_results}

\begin{figure}
\includegraphics[width=88mm]{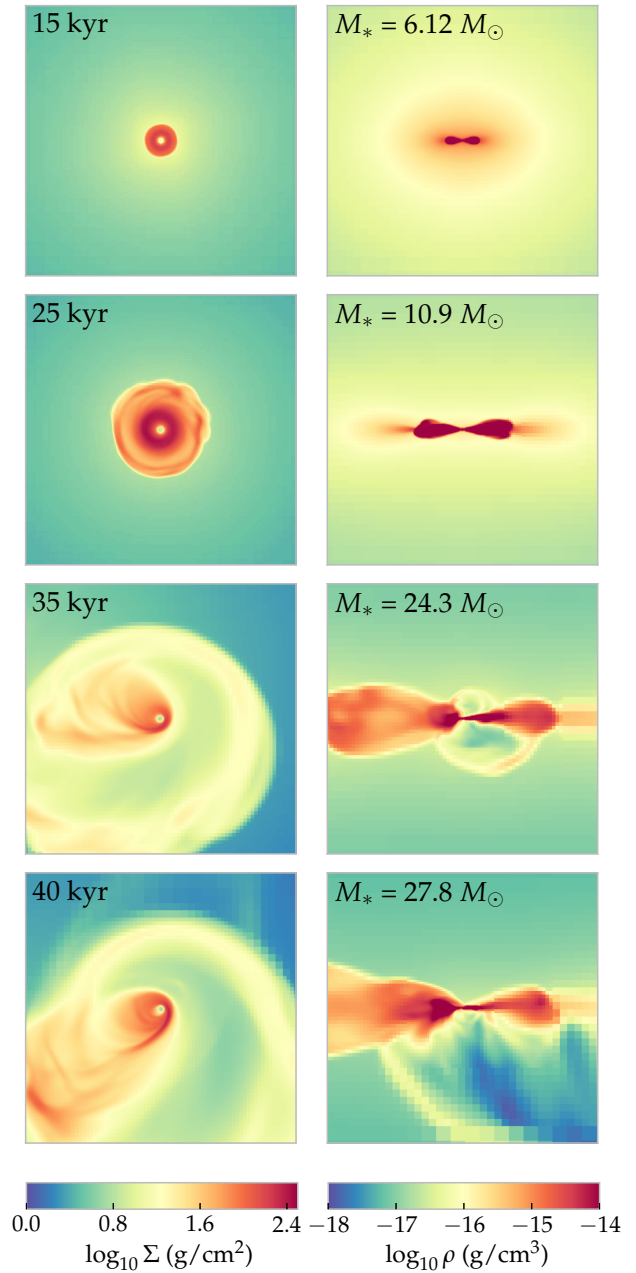}
\caption{A series of panels from our $100 \Msun$ simulation showing the time evolution of the disk in face-on and edge-on views. The times are indicated in the upper left corner of each row. The left column shows the projected gas density of the disk within a (3000 AU)$^2$ region. The colors indicating column densities are scaled from $\Sigma = 1$ g/cm$^2$ to $10^{2.5}$ g/cm$^2$. The right column is a slice of the volume density taken to show the edge-on view of the protostellar disk. The window is (3000 AU)$^2$ and the colors are scaled from $\rho = 10^{-18}$ g/cm$^3$ to $10^{-14}$ g/cm$^3$. The mass of the star in each pair of panels is indicated in the top-left of the second panel of each row.}
\label{fig:m100_evolution}
\end{figure}

The evolution of the protostellar core proceeds through approximately four stages: (1) collapse, (2) disk formation, (3) disk instability, and (4) radiation feedback. The gas is initially spherically symmetric in all our simulations and begins to contract gravitationally. Before long, a Jeans instability at the center of the volume results in the creation of a sink particle representing the gravitational collapse to stellar densities. The sink particle begins accreting material from the surrounding gas.

Figure \ref{fig:m100_evolution} shows snapshots taken from our 100 $\Msun$ simulation at approximately 15, 25, 35, and 40 kyr of evolution, as indicated. The left column shows the density projected along the $z$-axis, giving a face-on view of the protostellar disk. Each frame is centered on the location of the sink particle. The column densities in the first column are scaled from $\Sigma = 10^{0}$ g/cm$^2$ to $10^{2.5}$ g/cm$^2$. The column on the right shows the edge-on view, also centered on the sink particle, where a slice has been taken and volume densities within the slice plotted. The color of the indicated volume densities are scaled from $\rho = 10^{-18}$ g/cm$^3$ to $10^{-14}$ g/cm$^2$.

In this figure, each row shows the state of the simulation at the indicated time, zoomed in to a (3000 AU)$^2$ region around the sink particle. At 15 kyr (first row), the core has undergone gravitational collapse and a small protostellar disk has formed around the sink particle. In the second row, at 25 kyr, the disk is just beginning to undergo a gravitational instability. We see the formation of ripples at the edges of the disk, as well as a ring of dense material in the outer region of the disk. After a further 10 kyr of evolution, this disk shows clear spiral density waves (third row, Figure \ref{fig:m100_evolution}) and two radiatively-driven bubbles, one on either side of the accretion disk. The final row of Figure \ref{fig:m100_evolution} shows the end state of the simulation. The radiatively-driven bubble below the disk has continued to expand and is now more than 2000 AU across, while dense material flowing above the ray origin has momentarily quenched the direct radiation field above the disk and caused the bubble to disappear. In the face-on view, dramatic spiral density waves still swirl around the protostar, which continues to accrete material at a high rate ($\dot{M} \sim 10^{-3} \Msun$/yr).

When we compare this sequence of stages in each simulation, they all tell a similar story. In the 30 $\Msun$ core simulation, the star does not become quite luminous enough to drive bubble formation, but we still witness disk instability and the formation spiral waves. In the 200 $\Msun$ core simulation, the formation of star and disk proceed through the same stages, except the star is even more massive (reaching almost 44 $\Msun$) and so luminous that large outflow bubbles are formed on both sides of the disk and continue to grow until the end of the simulation.

We run each simulation until the timestep becomes too short to be practical. A strict upper limit on the simulation timestep is set by the gas velocities and the resolution. A highly luminous star will launch powerful outflows. These gas motions must be resolved. We observe velocities exceeding $10^7$ cm/s. With a grid resolution of $\sim 10^{14}$ cm, this gives an upper limit to the timestep of $\Delta t_{\textrm{sim}} \le 10^7$ seconds, which is shorter than a year.

At the time when we halted our simulations, each simulation had produced only a single sink particle. For an initial core mass of $30 \Msun$, the resulting sink attained a final mass of $5.48 \Msun$ after 81.4 kyr of evolution. The simulation that began with a core mass of $100 \Msun$ produced a star of $28.84 \Msun$, whereas our $200 \Msun$ simulation produced a star of $43.65 \Msun$.

The freefall time for a sphere of uniform density $\rho$ is given by
\begin{equation}\label{eqn:freefall_time}
t_{\textrm{ff}} = \sqrt{\frac{3 \pi}{32 G \rho}},
\end{equation}
where $G$ is Newton's constant. It represents the characteristic timescale for gravitational collapse. We estimate the freefall for our protostellar cores using the mean density. More massive cores have shorter freefall times, resulting in higher accretion rates onto the protostar. The simulations ran for 0.85, 0.79, and 0.59 freefall times for each of the initial 30 $\Msun$, 100 $\Msun$, and 200 $\Msun$ core simulations, respectively.

\begin{figure*}
\includegraphics[width=1.0\textwidth]{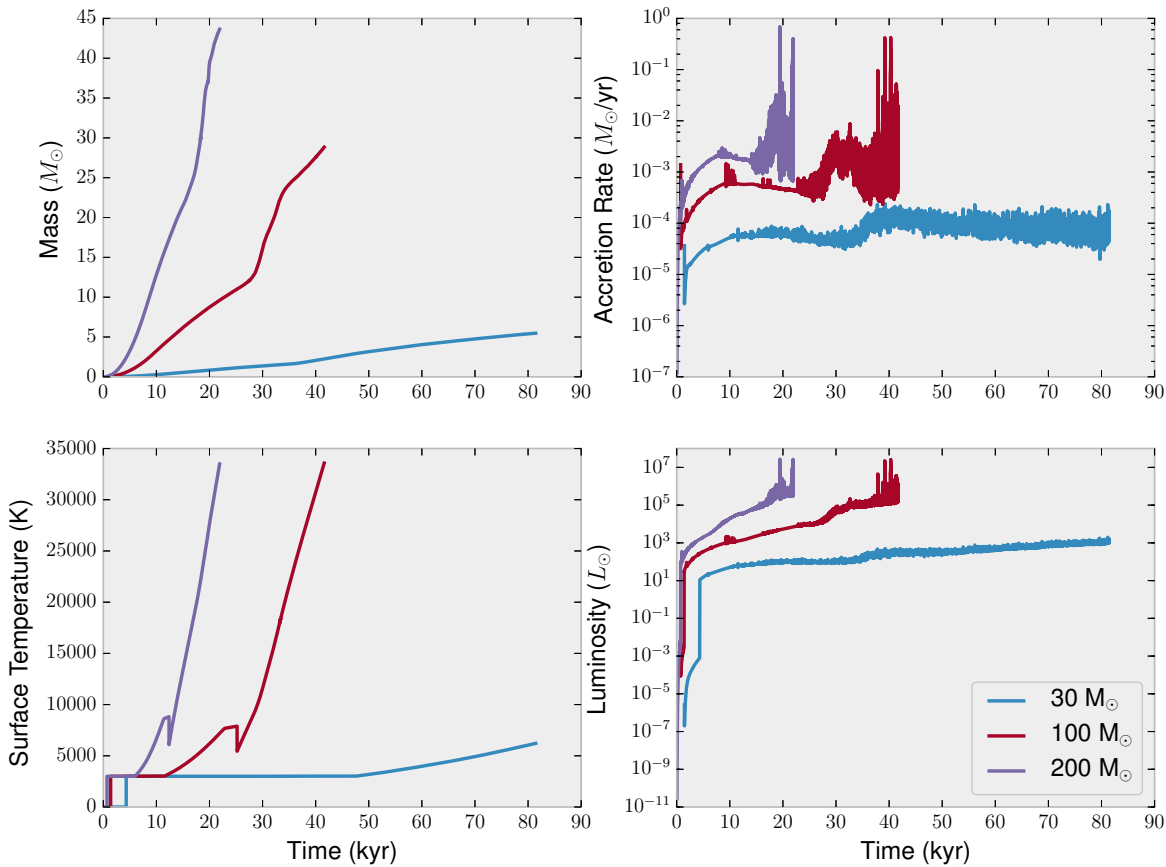}
\caption{The time-evolution of the sink particle formed in each of the three simulations (30, 100, and 200 $\Msun$ initial core mass). The sink particle represents the accreting protostar and its internal evolution is governed by a protostellar model. The top-left panel shows its mass. The top-right panel indicates the accretion rates onto the sink particles. The effective surfance temperature of the protostar is estimated by a protostellar model and is indicated in the bottom-left panel. The star radiates both due to the high accretion rate and its own intrinsic luminosity due to nuclear burning. The total luminosity of the sink is the sum of these two, and each sink particle's total luminosity is shown in the bottom-right panel.} 
\label{fig:sink_evolution}
\end{figure*}

We show the sink particle evolution in Figure \ref{fig:sink_evolution} as a function of time. In the upper left panel we show the mass histories of the star formed in each simulation. One interesting feature of these is that each possesses a `knee', where the rate of accretion shifts from one steady value to another relatively steady value. The knee appears to occur simultaneously with the disk instability, and we explore this below.

The accretion rate is shown in the upper right panel for Figure \ref{fig:sink_evolution} and shows the same transition to higher accretion rates. This occurs between 0.35 and 0.55 freefall times. The accretion rate increases by a factor of about 2--10, while also become more variable. We attribute both the increase in accretion rate and its increased variability to a transition in the accretion disk, which we discuss in detail below in section \ref{sec:disk_formation}.

The bottom two panels of Figure \ref{fig:sink_evolution} show the effective surface temperature (left) and luminosity (right) of the star, respectively. The temperature is given by a protostellar model that we introduced into the \FLASH code in \citet{Klassen+2012a} and is based on a one-zone model described in \citet{Offner+2009}. The drop in temperature observed in the 100 $\Msun$ and 200 $\Msun$ runs is on account of a swelling of the stellar radius at a time when the accreting protostar undergoes a change in its internal structure, switching from a convective to a radiative structure. The much larger stellar radius is initally cooler. The luminosity shown in the bottom right panel of Figure \ref{fig:sink_evolution} shows the combined stellar luminosity and accretion luminosity, $L_{\textrm{acc}} \propto GM_*\dot{M}_*/R_*$. This impact of protostellar evolution on the radiative feedback is in general agreement with earlier models of high-mass core collapse using 1D stellar evolution modeling (\citealp{KuiperYorke2013b}, but compare \citealp{HaemmerlePeters2016}).

We note that in each of our three main simulations, only a single sink particle is formed. This is in contrast to the similar calculation by \citet{Krumholz+2009}, which formed a binary companion with an orbital semi-major axis of 1280 AU. The secondary was about 1000 AU away from the central star, which is similar to the radius of the circumstellar accretion disks we form.

\begin{figure}
\includegraphics[width=88mm]{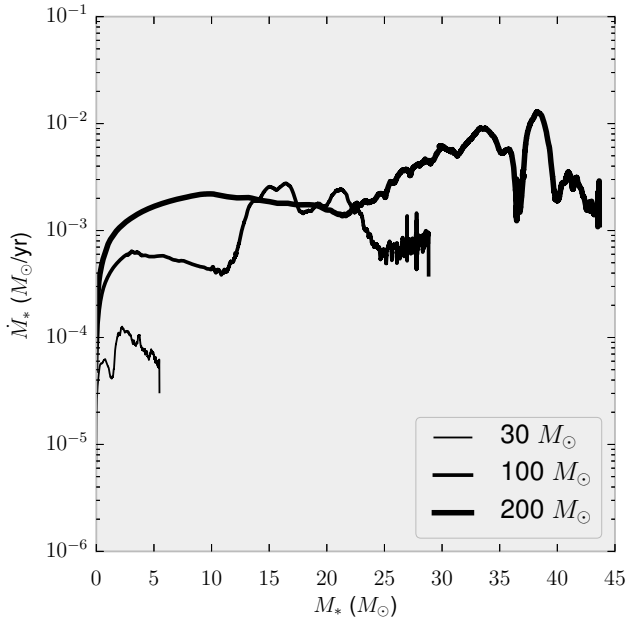}
\caption{The accretion rate plotted as a function of stellar mass for each of our three main simulations. The line thickness scales in order of ascending initial mass for the protostellar core.}
\label{fig:mdot_vs_m}
\end{figure}

In Figure \ref{fig:mdot_vs_m} we plot the evolution of the mass accretion rate as a function of the stellar mass for the sink particle in each of our three simulations. We apply a moving average filter of 1000 datapoints to smooth out the accretion rate, which is highly variable, especially at late times. We were unable to evolve the simulation to the point were accretion shuts down onto the sink particle. The figure is similar to Figure 11 in \citet{Kuiper+2010b}, but which featured simulation with axial and midplane symmetries. The 2D axisymmetric geometry of those simulations allowed for much longer runtimes. At late times accretion is seen to shut down, on account of both gas reservoir depletion and radiative feedback. We were unable to run our 3D non-axisymmetric simulations for the same duration, and so in Figure \ref{fig:mdot_vs_m} we do not yet see strong evidence of a shutdown in accretion and the gas reservoir is still far from fully depleted.

In the corresponding 3D simulation of \citet{Kuiper+2011}, the accretion disk also shows the formation of spiral arms and highly variable accretion rates, but no disk fragmentation. But the simulation could only be performed up to 12 kyr of evolution, hence, further evolution of the circumstellar disk, including binary formation, remains uncovered.

Before disk instability sets in, the accretion rate onto the star is governed more by spherically symmetric collapse of the initial core. \citet{Girichidis+2011} examined the collapse of a singular isothermal sphere \citep{Shu1977} and the scaling of accretion rate with the number of Jeans masses $N_J$ present in initial the protostellar core. We therefore compare the maximal accretion rate before disk instability sets in to the number of Jeans masses, and find that there is a linear relationship between accretion and $N_J$. The Jeans mass is given by
\begin{equation}
M_J = \frac{\pi^{5/2}}{6} \frac{c_s^3}{G^{3/2}\rho^{1/2}},
\end{equation} 
where $c_s$ is the isothermal sound speed and $\rho$ is the mean density of the spherical protostellar core. Thus, the number of Jeans masses present is
\begin{equation}
N_J = \frac{M_{\textrm{core}}}{M_J}
\end{equation} 

The linear scaling in accretion rate followed
\begin{equation}
\dot{M_*} = 9.638 \times 10^{-6} N_J \; \Msun \, \textrm{yr}^{-1}.
\end{equation}

The 30, 100, 200 $\Msun$ simulations contained 13.0, 79.3, and 224.4 Jeans masses, respectively, assuming isothermal 20 K gas. An approximately linear scaling is consistent with the theoretical results of \citet{Girichidis+2011}, although they were following the self-similar collapse of a singular isothermal sphere \citep{Shu1977} with power-law profiles $\rho \propto r^{-2}$.

\subsection{Disk formation and evolution}\label{sec:disk_formation}

\begin{figure}
\includegraphics[width=88mm]{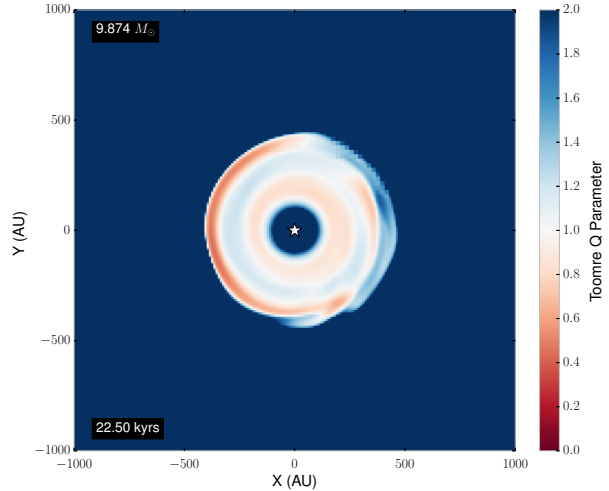}
\caption{The Toomre Q parameter, which quantifies the gravitational stability of disks. Values of $Q < 1$ are unstable to gravitational fragmentation and are colored in red. Stable regions are colored blue. Here we see the disk in our 100 $\Msun$ simulation becoming marginally unstable. The period of time immediately after this marks a strong increase in the accretion rate of material onto the star. The mass of the star is indicated in the top-left.}
\label{fig:toomreQ}
\end{figure}

\begin{figure*}
\includegraphics[width=1.0\textwidth]{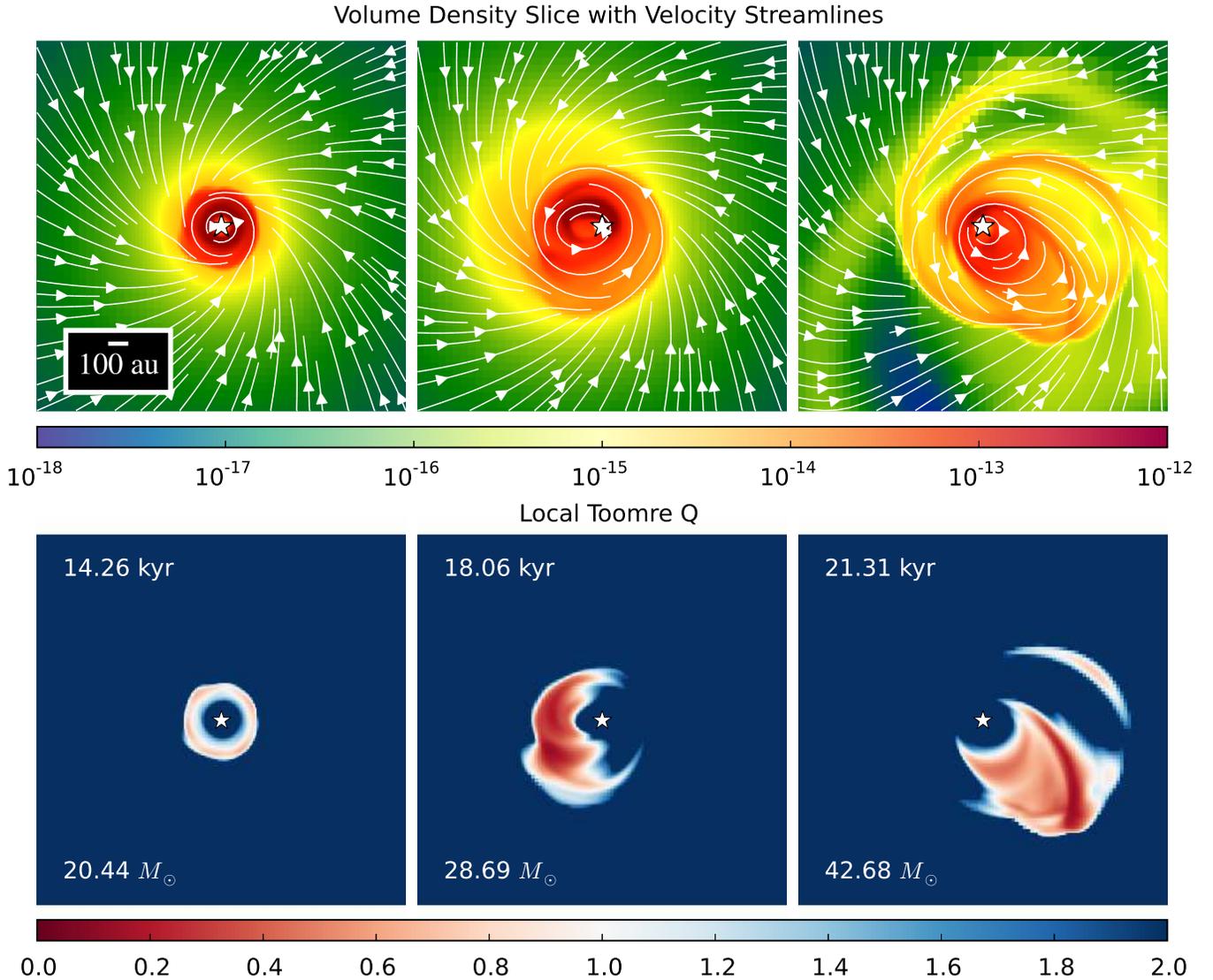}
\caption{Sequence of snapshots from the 200 $\Msun$ simulation showing face-on volume density slices (top row) and the local Toomre Q parameter (bottom row) in a (3000 AU)$^2$ region centered on the location of the sink particle. Scale bars have been added to show the volume density (in g/cm$^2$) and the value of the Q parameter (see Equation \ref{eqn:toomreQ}). Values of $Q < 1$ indicate disk instability. Velocity streamlines have been added to the volume density slice. The mass of the star is indicated in the bottom-left of each panel in the bottom row.}
\label{fig:streamlines+toomre}
\end{figure*}

The initial rotation results in the gradual formation of a disk around the sink particle. The disk assumes a flared profile and material continues to accrete onto both the sink particle and the disk. In Figure \ref{fig:sink_evolution}, the slower, steady accretion at the beginning of each simulation coincides with the growth phase of the disk around the sink particle.

\begin{figure*}
\includegraphics[width=1.0\textwidth]{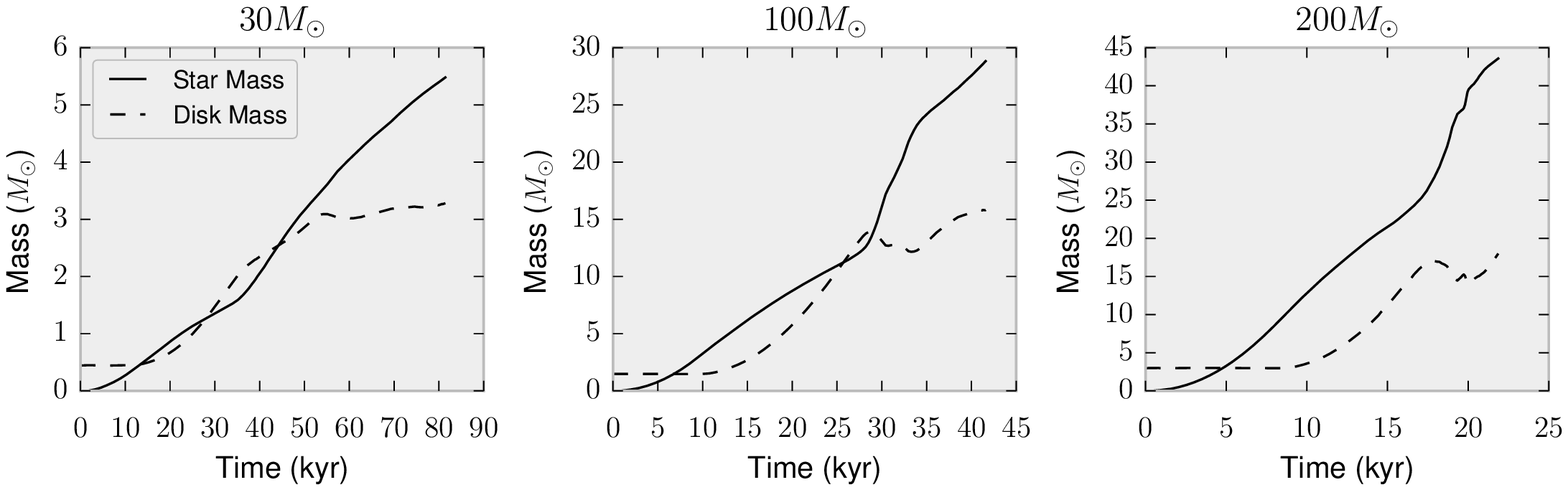}
\caption{The evolution of our three simulations, with initial core masses indicated above each frame. These show the simultaneous growth of the star and disk masses in each simulation.}
\label{fig:disk_vs_star_mass_evolution}
\end{figure*}

To measure the disk stability, we consider the Toomre Q parameter,
\begin{equation}\label{eqn:toomreQ}
Q = \frac{c_s \kappa}{\pi G \Sigma},
\end{equation}
where $c_s$ is the sound speed, $\kappa$ is the epicyclic frequency, $G$ is Newton's constant and $\Sigma$ is the column density. The Toomre stability criterion \citep{Toomre1964} predicts a disk instability for values of $Q < 1$. The epicyclic frequency $\kappa$ is equal to the angular velocity $\Omega$ for Keplerian disks. In the case of rigid rotation, $\kappa = 2\Omega$. The gas in our simulation starts out in rigid body rotation, but the disk then becomes Keplerian.

In Figure \ref{fig:toomreQ} we show the local Toomre Q value in our 100 $\Msun$ simulation at 22.5 kyr, the point in time just as the circumstellar accretion disk is becoming marginally unstable. Regions colored in blue show stable regions, while regions colored are unstable. White regions have a value for $Q \approx 1$, and are marginally stable. Within 5 kyr, the disk becomes highly asymmetric and the sink particle is perturbed from the center of the simulation volume. This event is accompanied by a marked increase in the accretion rate onto the sink particle, reaching values above $10^{-3} \Msun$/yr. The accretion rate is also far more variable during this phase of evolution, as is clearly visible in Figure \ref{fig:sink_evolution}. Analysis of the Toomre Q parameter in the other simulations tells the same story: material accretes onto the disk, trigging an eventual disk instability, resulting in increased accretion onto the star.

We show a sequence of snapshots from the 200 $\Msun$ simulaion in Figure \ref{fig:streamlines+toomre}. This time, we have paired a measurement of the local Toomre Q parameter (bottom row) with a volume density slice through the midplane of the disk (top row). The snapshots were taken at the same times, with the times indicated. The panels have been centered on the sink particle at show a (3000 AU)$^2$ region. In the volume density slices we have overplotted velocity streamlines that indicate the flow of material onto and through the circumstellar disk. We see that the flow merges with the spiral arms and that the spiral arms appear to act as accretion channels for material.

In Figure \ref{fig:disk_vs_star_mass_evolution}, we show the simultaneous growth of the star and disk masses in each of our three simulations. The disk mass is measured by considering a cylindrical volume, centered on the sink particle, and measuring the total gas mass contained within this cylinder. We choose a radius of 1000 AU, and a total height of 1000 AU. The resulting volume is large enough to capture the main extent of the disk, including approximately 2 pressure scale heights. The final disk masses of our three simulations were 3.3, 15.8, and 18.0 $\Msun$ for our 30, 100, and 200 $\Msun$ simulations, respectively.

In the $100 \Msun$ and $200 \Msun$ simulations, the `knee' in the stellar mass evolution, corresponds to an instability of the disk. It loses its axisymmetry and forms large spiral arms. At this time, Figure \ref{fig:disk_vs_star_mass_evolution} shows that the disk mass ceases to grow monotonically and we measure a temporary decrease in the overall disk mass. This may, however, be due to spiral arms flinging material outside the 1000 AU radius of the cylinder we are using to measure the disk mass. 

The final disk masses were measured at 3.3 $\Msun$, 15.8 $\Msun$, and 18.0 $\Msun$ for our 30 $\Msun$, 100 $\Msun$, and 200 $\Msun$ simulations, respectively. The latter two are relatively close in mass, but the disk accretes faster in the $200 \Msun$ case. It also has the highest time-averaged star mass to disk mass ratio (1.96). The 30 $\Msun$ simulation had the lowest time-averaged star-to-disk mass ratio (1.07), with the 100 $\Msun$ falling in between with 1.45.

\subsection{Disk Fragmentation}

\begin{figure*}
\includegraphics[width=\textwidth]{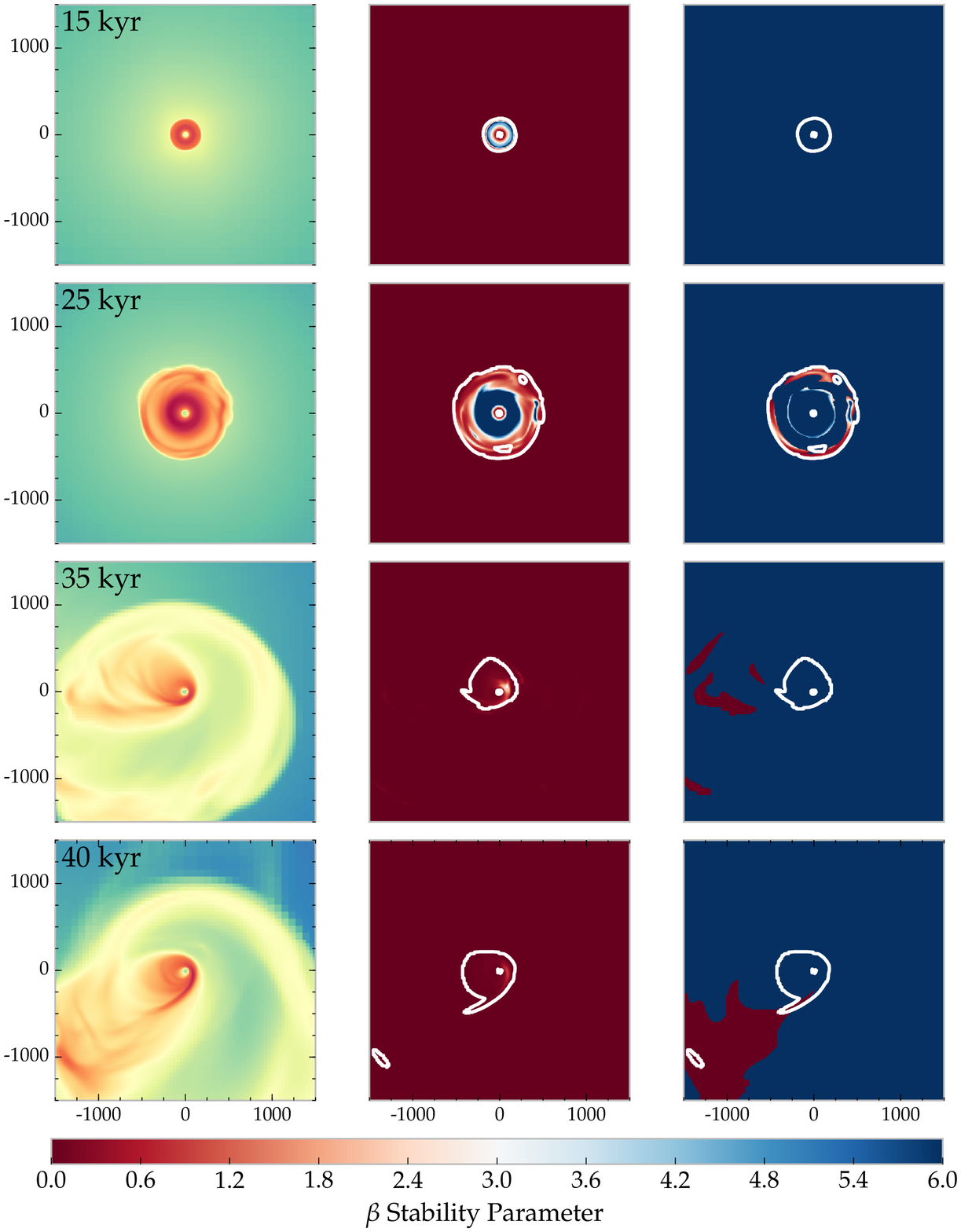}
\caption{An analysis of the local stability of the entire protostellar disk from the 100 $\Msun$ simulation. The rows show the state of the simulation at the stated times, which are identical to the ones shown in Figure \ref{fig:m100_evolution}. The left column shows the column density of the disk. The center column shows the $\beta$ stability parameter for the entire disk. The right column shows the $\beta$ stability parameter again, but with Toomre-stable regions ($Q > 1$) masked out. Colored in red, therefore, are only those regions that are both Toomre-unstable and $\beta$-unstable. These are expected to be gravitationally unstable. The white contour in the panels of the middle and right columns indicate those regions where the density exceeds the threshold for sink particle formation, $\rho_{\textrm{thresh}} = 1.4 \times 10^{-14}$ g/cm$^3$.}
\label{fig:gammie_beta}
\end{figure*}

Contrary to a very similar calculation performed by \citet{Krumholz+2009}, we do not form a binary star as the disk goes unstable, or any additional sink particles. Despite the instability of the disk, fragmentation is not seen. This is a significant result given that spectroscopic surveys show a high binary fraction among massive stars. \citet{Chini+2012} performed a high-resolution radial velocity spectroscopic survey of massive stars within the Milky Way galaxies that included about 250 O-type stars and 540 B-type stars. Their results indicated that over 82\% of stars with masses above 16 $\Msun$ form close binaries. This fraction drops precipitously for lower-mass stars.

How many stars are expected to form per protostellar core? \citet{GoodwinKroupa2005} argue from dynamical constraints and observations that protostellar cores should produce only 2 or 3 stars. This contrasts with some numerical simulations, in which the protostellar core fragments into a greater number of stars, as in \citet{Goodwin+2004} or the simulations of \citet{Krumholz+2009}. It also contrasts with our simulations, which show a massive disk that does not fragment into more stars.

We examine several approaches to the question of fragmentation. First, we follow the analysis of \citet{RogersWadsley2012}, who studied the fragmentation of protostellar disks and derived a Hill criterion for the spiral arms \citep[although, see][]{Takahashi+2016}. These are the sites most likely to begin fragmenting. They discovered that if the width of these spiral arms is less than twice the Hill radius, then fragmentation of the spiral arms is expected because the self-gravity of that spiral arm segment dominates the tidal forces from the star (manifested as rotational shear).

To quantify this, we look at the Hill radius, as defined in \citet{RogersWadsley2012},
\begin{equation}\label{eqn:hill_radius}
H_{\textrm{Hill}} = \left[\frac{G\Sigma l^2}{3\Omega^2}\right]^{1/3},
\end{equation}
where $G$ is Newton's constant, $\Sigma$ is the column density of the spiral arms segment, $l$ is the width (thickness) of the spiral arm, $\Omega$ is the angular velocity, assuming a Keplerian disk. Here $\Sigma l^2$ is a measure of the mass contained within the Hill sphere.

The Hill criterion for stability against fragmentation is
\begin{equation}\label{eqn:hill_criterion}
\frac{l}{2H_{\textrm{Hill}}} > 1.
\end{equation}

\citet{RogersWadsley2012} demonstrate the validity of this criterion through hydrodynamic simulations of protostellar disks, finding gravitational fragmentation occurring when $l/(2H_{\textrm{Hill}}) < 1$. We test whether the spiral arms present in our own calculation are indeed stable by looking at snapshots from our simulations at an evolved state. We then took a cross-section of the spiral arm where column densities were greatest and measured the mass along the cross section and the angular rotation rate. In each case the spiral arms were 100 AU wide, and the masses through a section of the spiral arms at the cross section were 0.033 $\Msun$, 0.15 $\Msun$, and 0.29 $\Msun$ for the 30, 100, and 200 $\Msun$ simulations, respectively. Meanwhile, the rotation rates at these locations were $\Omega = 3\times10^{-10}$, $6\times10^{-10}$, and $6\times10^{-10}$, respectively. Finally, this resulted in Hill criterion values of $l/(2H_{\textrm{Hill}}) = 2.95$, 2.82, 2.26, respectively, consistent with the lack of fragmentation observed in our simulations. 

This analysis nevertheless focuses on the role of shear stabilization in spiral arms. We now turn to the role of cooling and disk fragmentation across the entire disk. \citet{Gammie2001} and \citet{JohnsonGammie2003} investigated nonlinear gravitational instability in numerical models of thin, Keplerian disks. By considering the cooling time $\tau_c$ and the angular rotation rate $\Omega$, a stability parameter can be defined,
\begin{equation}\label{eqn:gammie_beta}
\beta = \tau_c \Omega,
\end{equation}
which must be greater than some critical value in order for the disk to remain gravitationally stable. Gas with short cooling times relative to the orbital period is expected to cool rapidly---opening the way to fragmentation if the gas is also sufficiently self-gravitating. The critical value for $\beta$ is established via numerical simulations, but depends critically on which heating and cooling mechanisms are present in the simulation. The cooling time can be defined as the internal energy of the gas, divided by its cooling rate. For optically thick disks, \citet{JohnsonGammie2003} found fragmentation occurs for values of $\beta = \langle\tau_c\rangle \Omega \sim 1$, where $\langle\tau_c\rangle$ is the disk-averaged cooling time. It is important to note that the $\beta$-criterion for stability does not replace the Toomre Q stability criterion, but rather complements it. That is, a disk must be both Toomre-unstable and have a Gammie $\beta$ less than some critical value for fragmentation to occur.

We choose to look at the local cooling rate. Since we possess information about the rate of radiative flux loss from the disk, we define the cooling time as
\begin{equation}\label{eqn:cooling_time}
\tau_c = \frac{E}{\langle\left| \vec{F}_{\textrm{rad,z}}\right|\rangle},
\end{equation}
where $E$ is the column internal energy integrated through the disk. Meanwhile, $\langle\left| \vec{F}_{\textrm{rad,z}}\right|\rangle$ is the mean radiative flux in the vertical direction (away from the disk). Our cooling time, therefore, is a direct function of the mean density, temperature, and opacity of a vertical column through the circumstellar disk. Since we know the rotation rate at each point, we then calculate a local $\beta = \tau_c \Omega$.

In Figure \ref{fig:gammie_beta} we compare side-by-side the column density and disk stability for the circumstellar disk in our 100 $\Msun$ simulation, taking snapshots from the simulation at the same times as in Figure \ref{fig:m100_evolution}. As in that earlier figure, we show the column density in the left column. The middle column shows the value for $\beta$ throughout the disk. We see that the cooling time is short relative to the orbital period and the disk should be prone to rapid cooling. Recall that \citet{JohnsonGammie2003} found that fragmentation occurs for values of $\beta \sim 1$.

However, if we consider only those regions which are \textit{also} Toomre-unstable, as we do in the right column of Figure \ref{fig:gammie_beta}, we see that most parts of the disk remain stable. In generating the panels for this column, we created an image mask, i.e. we filter the pixels based on value of the Toomre Q parameter at that location and then draw the $\beta$ parameter. Regions where $Q > 1$ are colored uniformly in blue; these regions are stable regardless of the value of $\beta$. Additionally, we draw a white contour in the panels of the middle and right columns to indicate those regions where the density exceeds the threshold for sink particle formation, $\rho_{\textrm{thresh}} = 1.4 \times 10^{-14}$ g/cm$^3$.

For our $100 \Msun$ simulation, the threshold density for sink particles is $\rho_{\textrm{thresh}} = 1.4 \times 10^{-14}$ g/cm$^3$. We draw a white contour in Figure \ref{fig:gammie_beta} to indicate those regions within a disk slice possessing densities greather than this threshold value. 

Taken together, the disk is largely stable across much of its extent. Nevertheless, Figure \ref{fig:gammie_beta} shows the disk is not completely stable everywhere, at least predicted by linear stability analysis. This can be seen especially in the final set of panels at 40 kyr. There appears an island of high density gas that is both Toomre-unstable and $\beta$-unstable at about 1200 AU separation from the central star. This is a candidate region for collapse. Given more time it could indeed collapse to form a single wide binary companion to the central star. As pointed out in \citet{Takahashi+2016}, fragmentation is a nonlinear outcome of gravitational instability and highly dependent on initial conditions. They demonstrate that the only necesssary condition for the formation of spiral arms is that $Q < 1$, and the only necessary condition for the fragmentation of these arms is $Q < 0.6$. Taken together with our stricter sink particle criteria, as described in Section \ref{sec:sink_particles}, this accounts for the lack of secondary fragmentation over the timescales simulated.

The gravitational collapse that formed the first star in our simulation, and the strong radiative accelerations produced by the intense luminosity of that massive star, conspire with the Courant condition to strongly limit the timestep size of our simulation. The timestep is now so small that the simulation has now effectively stalled.

\subsection{Radiatively-Driven Bubbles}

In the vicinity of a massive star, the radiative force can exceed the force of its gravitational attraction, resulting in radiatively-driven winds or bubbles, and possibly halting any further accretion. Radiation pressure is also one of the main mechansism for the disruption of giant molecular clouds (GMCs) and the cumulative radiation pressure from star clusters may drive large-scale galactic outflows \citep{Murray+2010,Murray+2011}. 

The Eddington luminosity describes the force balance between radiation and gravity:
\begin{equation}
L_{\textrm{edd}} = \frac{4 \pi G M_* c}{\kappa},
\end{equation}
where $\kappa$ denotes the opacity of the absorbing medium. We estimate the direct radiation pressure using temperature-dependent gray opacity \citep[see][Figure 1]{Klassen+2014} for the dust. The body force on the dust grains is given by
\begin{equation}
f_{\textrm{rad}} = \rho \kappa_P\left(T_{\textrm{eff}}\right)\frac{F_*}{c},
\end{equation}
where $\kappa_P = \kappa_P(T_{\textrm{eff}})$ is the Planck mean opacity, $T_{\textrm{eff}}$ is the effective temperature of the star, estimated by protostellar model, and $F_*$ is the stellar flux. 

\begin{figure*}
\includegraphics[width=\textwidth]{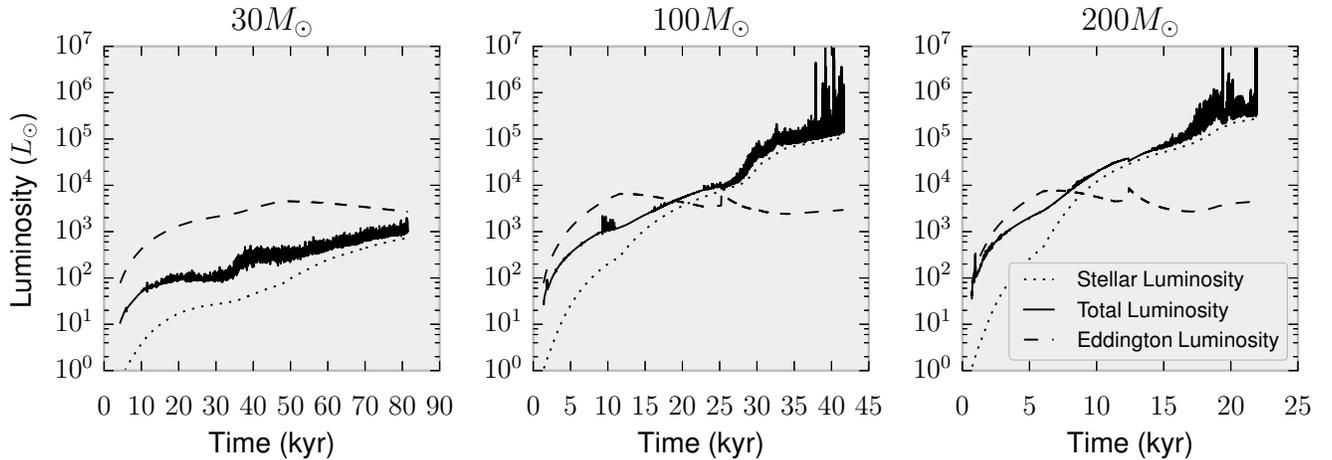}
\caption{The evolution as a function of time of the total luminosity (solid line) of the star formed in each of our three main simulations. In each case it is compared to the star's Eddington luminosity (dashed line), the threshold where radiation forces exceed gravitational forces. In the $100 \Msun$ and $200 \Msun$ simulations, the star's luminosity becomes super-Eddington. It is in these simulations that radiatively-driven bubbles are observed.}
\label{fig:luminosity_evolution}
\end{figure*}

In Figure \ref{fig:luminosity_evolution}, we show the evolution in each of our three simulations the stellar radiative flux. It is plotted here as a function time and compared, in each case, to the star's Eddington luminosity. We see that only in the $100 \Msun$ and $200 \Msun$ simulations does the star become bright enough to exceed the Eddington limit. The excess radiation pressure in these two more massive simulations results in the formation of radiatively-driven outflow bubbles that expand away from the star above and below the accretion disk, sweeping up a shell of material and leaving a low-density gas in their wake. Radiation spills into these more optically thin regions via a kind of `flashlight effect' \citep{Nakano1989,YorkeBodenheimer1999,YorkeSonnhalter2002}, helping to drive the outward expansion of the bubble. The flashlight effect as described in \citet{YorkeBodenheimer1999} is the anisotropic distribution of radiative flux around a star after the formation of a circumstellar disk. The disk in effect channels the radiation along the polar axis, creating a `flashlight'.

We note also that in the low-mass $30 \Msun$ simulation, the star's luminosity remains sub-Eddington, but continues to creep steadily upward. At the time that we halted the simulation, the star's accretion rate, though not exceptionally high ($10^{-4} \Msun$/yr), was showing no signs of slowing down (see Figure \ref{fig:sink_evolution}). At this rate, the star might yet have become radiant enough to drive an outflow bubble.

\begin{figure}
\includegraphics[width=88mm]{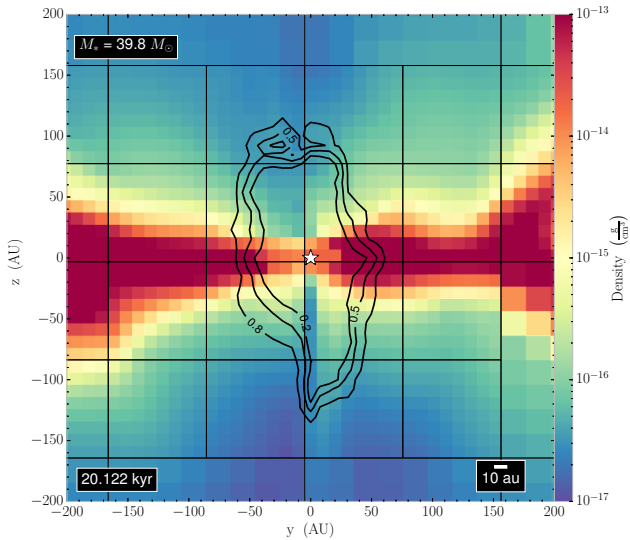}
\caption{Volume density slice of a (400 AU)$^2$ region through our 200 $\Msun$ simulation, showing an edge-on view of the circumstellar disk. The slice is centered on the sink particle, representing a 39.8 $\Msun$ star. \textsc{flash}'s block structure is shown in the grid, with each block containing $8^3$ cells. Overplotted are contours showing dust fraction. Nearest the star, the dust is completely sublimated. Contours show dust sublimation correction factors of 0.2, 0.5, and 0.8.}
\label{fig:dust_evap}
\end{figure}

Based purely on the above considerations, it might be surprising that accretion continues unabated even after the star goes super-Eddington. However, the above estimation of the Eddington luminosity did not account for dust sublimation, which occurs at the inner edge of the disk. In our simulations, the dust sublimation was approximated by Eqs.~(\ref{eqn:dust_evap_temperature}) and (\ref{eqn:dust_evap_correction}), and this results in a low opacity in regions where the gas has been heated to high temperature. Radiative forces can no longer couple to the gas via the dust, and the gas continues to move inward and accrete onto the star. In \citet{Kuiper+2010b,KuiperYorke2013a} it was shown that the inclusion of the dust sublimation front was a requirement for continued accretion via the protostellar disk and enhanced the anisotropy of the radiation field, contributing to the flashlight effect. Our resolution (10 AU) is not quite as high as those (1.27 AU radially at the inner edge of the disk) in \citet{Kuiper+2010b}, but we also show continued disk accretion and a strong flashlight effect.

Figure \ref{fig:dust_evap} shows the dust sublimation front at a snapshot in time from our most massive (200 $\Msun$) core simulation, after about 20 kyr of evolution. We plot an edge-on volume density slice intersecting the sink particle and circumstellar disk and showing a region 400 AU wide on a side. The \FLASH block-grid structure is overplotted for reference. Each block contains $8^3$ cells. The sink particle, representing a $39.8 \Msun$ star, is located at the center of the frame and is in the process of driving a radiative bubble. The star has grown to almost 20\% of the initial mass of the protostellar core. Figure \ref{fig:dust_evap} shows contours of dust fraction overplotted. A value of 0.0 implies total dust destruction and a value of 1.0 implies no sublimation. We show contours at values of 0.2, 0.5, and 0.8. Dust sublimation regions are also optically thin, and Figure \ref{fig:dust_evap} shows clearly how radiation is channeled in the polar direction.

\begin{figure}
\includegraphics[width=88mm]{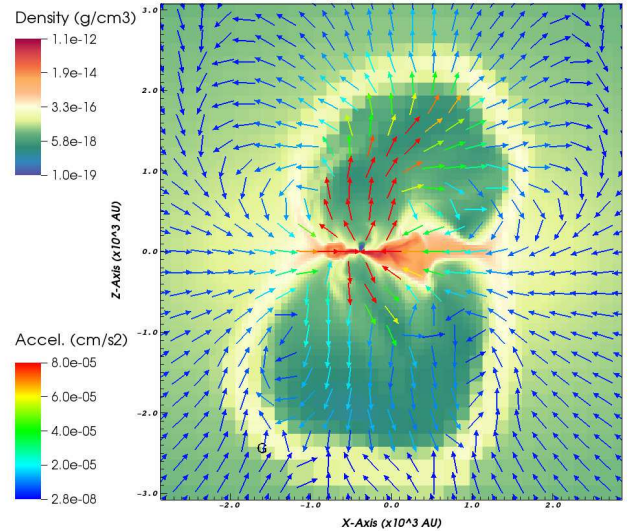}
\caption{Acceleration vectors showing the net acceleration (in units of cm/s$^{-2}$) of the gas due to radiative forces and gravity (both the gravity of the star and the gas). Vectors are overplotted on a background indicating volume density slice of an approximately (6000 AU)$^2$ region through our 200 $\Msun$ simulation after about 21.8 kyr of evolution, showing an edge-on view of the circumstellar disk and radiatively-driven bubble. The slice is centered on the sink particle, representing a 43.5 $\Msun$ star.} 
\label{fig:net_accelerations}
\end{figure}

In the classical Rayleigh-Taylor instability, a heavy fluid cannot be stably supported by a light fluid. In Figure \ref{fig:net_accelerations}, we plot the net acceleration based on the radiative and gravitational forces present. The data is taken from our 200 $\Msun$ simulation, after 21.8 kyr of evolution. The radiative acceleration includes both the direct and the diffuse radiation field, while the gravitational acceleration include the attractions of both the star and the self-gravity of the gas. The figure shows the volume density of the gas in a slice centered on the sink particle so that an edge-on view of the disk is shown. Overplotted are the acceleration vectors with magnitude coded in the color. The direct radiation field exerts very powerful accelerations immediately surrounding the star. This energy is absorbed and reprocessed so that the diffuse radiation field dominates the radiative accelerations in shadowed regions, the disk, and far away from the star.

The star has reached a mass of 43.5 $\Msun$ and has driven a radiative bubble above and below the disk. While along the edge of the lower bubble wall, there is a null surface in the acceleration---the radiative and gravitational forces roughly cancel, though momentum is likely still causing the bubble to expand. On the upper edge, however, we see that a net outward acceleration extends well beyond the bubble wall, ensuring that the bubble will continue to expand outward. It also means that the classic condition for the Rayleigh-Taylor instability (that of a heavy fluid supported by a light fluid) is not met. The acceleration vector is outwards, so that the ``lighter'' fluid (FLD approximates the radiation field as a fluid) sits atop the heaver fluid of the bubble wall. This is another reason why our bubbles appear roughly uniform and show no signs of the Rayleigh-Taylor instability. We checked the bubble in the 100 $\Msun$ simulation and there, too, there was net outward acceleration. We therefore conclude that Rayleigh-Taylor instabilities are absent in these early phases, which are instead dominated by net outward acceleration driven by the radiation field.

\begin{figure*}
\includegraphics[width=\textwidth]{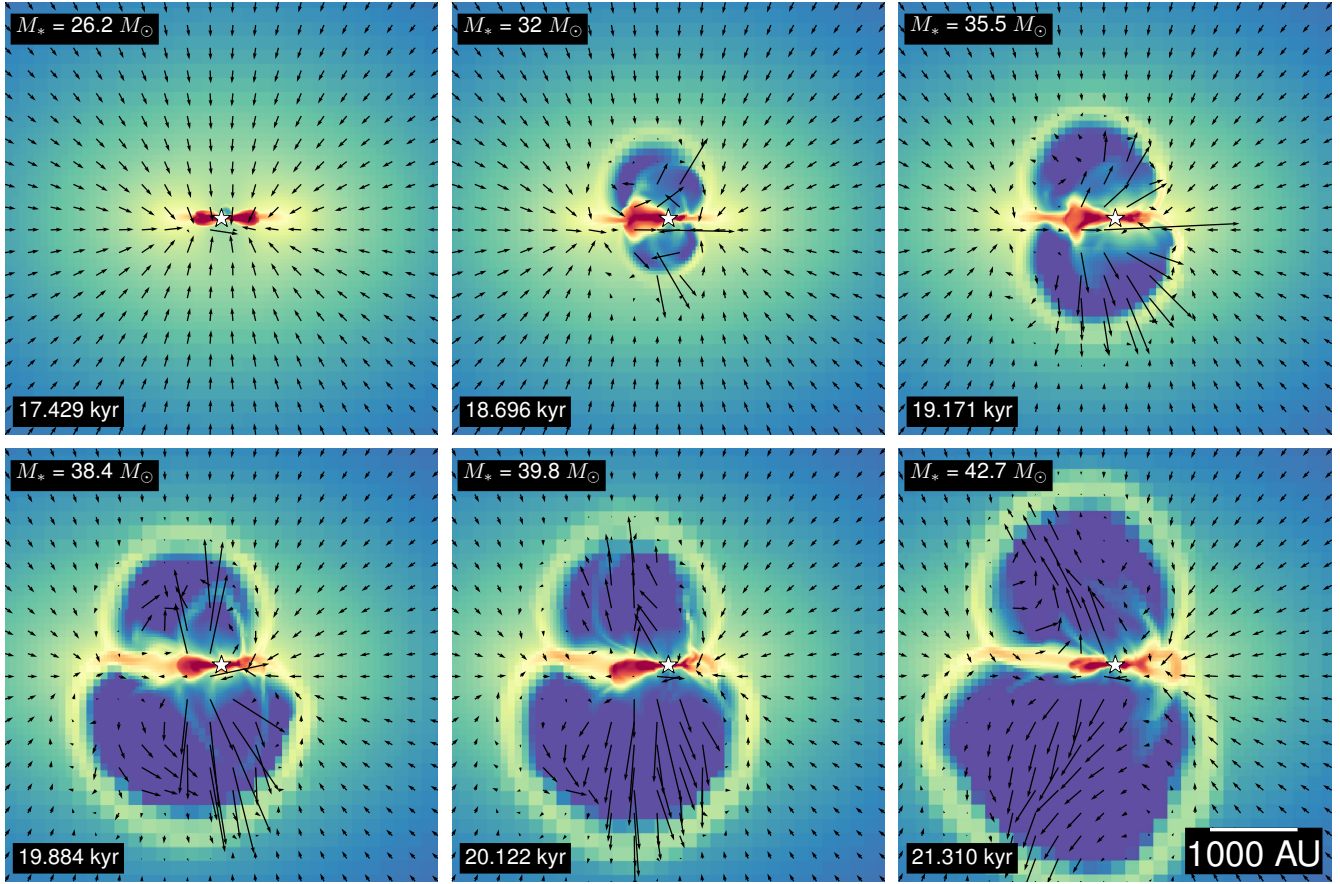}
\caption{Edge-on volume density slices, centered on the sink particle, showing in a (5000 AU)$^2$ region the evolution of a radiatively-driven bubble in our 200 $\Msun$ simulation. Shown are six snapshots at different points in the simulation with the times indicated. Velocity vectors have been overplotted. Volume densities have been scaled from $\rho = 10^{-17}$ g/cm$^3$ to $10^{-13}$ g/cm$^3$. The mass of the star is indicated in the top-left of each panel.}
\label{fig:bubble_evolution}
\end{figure*}

In \citet{Krumholz+2009}, the Rayleigh-Taylor instability that was observed in their simulations served to feed material back onto the disk and the star. If the disk can be fed continually, either from the outer regions of the simulation, or from material escaping the polar outflow, then the star should be able to continue accreting, provided the thermal radiation pressure within the disk itself is not so strong as to reverse the accretion flow. In contrast, \citet{Kuiper+2012} modeled the high-mass pre-stellar core collapse comparing two different schemes for radiative feedback, namely the hybrid scheme used also in this investigation and the FLD-only scheme used in \citet{Krumholz+2009}; as a result, only simulations using the FLD approximation yield a radiative Rayleigh-Taylor instability, simulations with the hybrid scheme show stable outflows for stars in the super-Eddington regime. \citet{Kuiper+2010b} showed that the anisotropy of the radiation field prevents the thermal radiation from ever becoming strong enough to halt disk accretion. It then becomes a question as to whether the disk can be continually fed.

Figure \ref{fig:bubble_evolution} clearly shows that collapsing gas is deflected along the walls of the bubble, and flows onto the disk. Here we look at the gas density and velocities in a series of slices from our $200 \Msun$ simulation. We centered individual slices on the position of the sink particle and took an edge-on view of a (5000 AU)$^2$ region. The slice intersects the circumstellar accretion disk and shows a radiatively-driven bubble forming. The second-to-last panel was taken at the same time as Figure \ref{fig:dust_evap}, which is also near the end of our simulation. The star's luminosity is super-Eddington in each of the panels shown (compare Figure \ref{fig:luminosity_evolution}) and the outflow bubbles develop into uniformly, both above and below the disk, and grow to be over 2000 AU in diameter. Interior to these bubbles is evidence of smaller bubbles, expanding in successive waves and finally bursting their walls.

We overplot velocity vectors to show the gas motion. Interior to the radiatively-driven bubbles, the gas is being driven away from the star asymmetrically. The direct radiation field from the star is highly sensitive to the gas distribution: it is strongly attenuated in the disk plane, whereas the polar direction is more optically thin. The result is rarified gas being accelerated to high velocity along the polar axis. The highest measured velocity can be over 60 km/s. There is no evidence of Rayleigh-Taylor instability.

Outside of the radiatively-driven bubble, we see gas continuing to fall gravitationally towards the star. This motion is deflected at the bubble wall and we observe gas moving along the bubble walls and onto the accretion disk. Material simply flows along the bubble wall and in this way continues to supply the protostellar disk with matter. We note that the details may depend on opacity effects: while our work uses gray atmospheres, long-wavelength radiation was seen to accelerate matter outside the cavity walls in multifrequency simulations by \citet{Kuiper+2010b,Kuiper+2011,Kuiper+2012}.

The motion of the star relative to its massive, asymmetric disk results in the star becoming periodically buried in disk material. This has the result of temporarily reducing the direct irradiation of the polar cavities. It is also the likely reason for the formation of successive shells of outward-moving material visible in Figure \ref{fig:bubble_evolution}. It also contributes to the sometimes asymmetric appearance of the radiatively-driven bubbles, as seen in Figure \ref{fig:m100_evolution} for the 100 $\Msun$ simulation.

Given the symmetric nature of the initial conditions of the simulation, one might expect the formation of radiatively-driven bubbles with North-South symmetry. The Cartesian grid structure easily introduces small numerical perturbations that break axial symmetry and trigger gravitational instability and spiral wave formation, but these do not explain the breaking of plane symmetry. \citet{Pringle1996} showed analytically that even initially flat disks with a central radiation source are subject to a warping instability caused by radiative torques. Also, small numerical perturbations can also be introduced by the diffusion solver that iterates over grid cells until specific global convergence criteria are met. If the solution is converged, then the current iteration ceases without needing to visit the remaining grid cells. This then breaks North-South symmetry. Eventually, differential shadowing in each hemisphere by material spilling onto the star then results in the asymmetric bubbles as seen in our simulation.

\subsection{Accretion Flows and Outflows}\label{sec:accretion_flows}

Finally, we analyzed the accretion flows in our simulation at various times. To do this, we found it very helpful to create a type of graph profiling the material moving towards or away from our sink particle as a function of the polar angle as measured from the `north' vector at the location of the star.

We used {\tt yt}'s profiling tools to examine the radial velocity at the location of the sink particle, being careful to subtract the sink particle's own motion from the gas velocity. We then produced an azimuthally-averaged gas velocity profile as a function of polar angle. 

\begin{figure*}
\includegraphics[width=\textwidth]{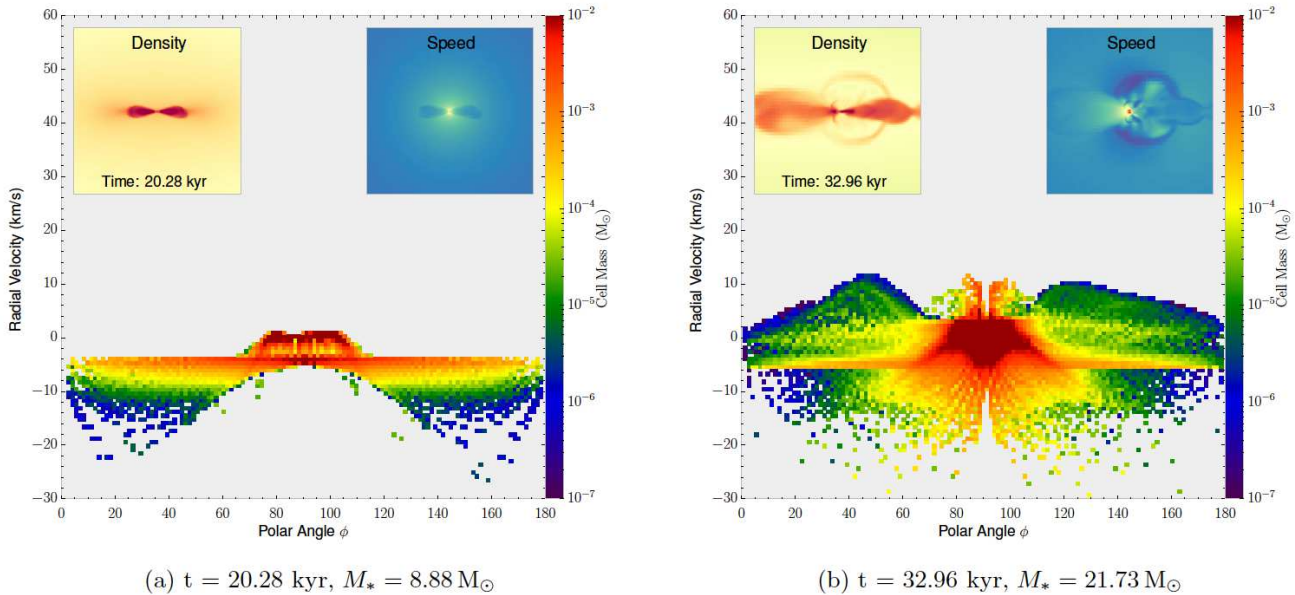}
\caption{Azimuthal accretion and outflow profile at two different times during the $100 \Msun$ simulation within a 1000 AU radius centered on the sink particle. In each of these we measure the azimuthally-averaged radial velocity of the gas relative to the star. This gives a picture of how much gas is moving towards or away from the star as a function of the polar angle. The first panel shows the simulation after the formation of a flared protostellar disk. The disk is even and hasn't yet gone unstable. The second frame shows the state of the simulation after the star has gone super-Eddington and a radiatively-driven bubble has formed. Gas is accelerated to around 10 km/s away from the star in the regions above ($20^{\circ}$--$70^{\circ}$) and below ($110^{\circ}$--$160^{\circ}$) the disk.}\label{fig:m100_accretion_profile}
\end{figure*}

We show two examples of this in Figure \ref{fig:m100_accretion_profile}, where we look at gas motions around the star formed in our $100 \Msun$ simulation. The first panel shows the state of the simulation after 20.28 kyr. The star's mass is about $8.9 \Msun$. The panel has two inset frames, one showing a representation of gas density, the other the magnitude of the gas velocity, both within an edge-on slice centered on the star. Colorbars have been omitted for the inset frames so as not to overly complicate the figure. We profile the gas motions within a spherical volume, centered on the sink particle, with a radius of 1000 AU. The inset frames show a slice through this volume, i.e.~their width of 2000 AU matches the diameter of the profiling volume.

The main panel has been colored like a two-dimensional histograph, where the color indicates the amount of gas (in units of solar masses) measured at a particular radial velocity and along a particular polar angle, while averaging along the azimuthal angle. In the left panel of Figure \ref{fig:m100_accretion_profile}, we see steady inward gas motion across virtually all angles. This can be seen in the horizontal line at $v_r = - 4$ km/s. Some of the fastest inward gas motion happens along angles $\pm 30^{\circ}$ from the rotation axis. This is seen in the gas components with radial velocities from $v_r = - 10$ to $-20$ km/s.

At a polar angle of $\sim 90^{\circ}$ (the plane of the protostellar disk) we see a dense gas component with a radial velocity of $v_r \sim 0$ km/s. This is the gas that is within the accretion disk and is orbiting the star in the plane of the disk. We observe that this radially stationary component is spread over a range of angles, from about $75^{\circ}$ to about $105^{\circ}$.

Within the disk plane is another velocity component that indicates gas moving inward. At $\phi = 90^{\circ} \pm 5^{\circ}$ is some of the highest density gas, moving with a velocity of about $v_r = - 4$ km/s. This we understand to be disk accretion. At this time in our $100 \Msun$ simulation, the star's accretion rate is $\dot{M} \approx 3 \times 10^{-4} \Msun$/yr.

In the second panel of Figure \ref{fig:m100_accretion_profile}, the $100 \Msun$ simulation is shown at a later time, at almost 33 kyr of evolution. The star has grown to about $21.7 \Msun$. The disk has already gone Toomre-unstable and is no longer axisymmetric. Gas orbiting the star inside the accretion disk moves in elliptical orbits and the measured accretion rates onto the particle are much higher ($\dot{M} \approx 1$--$2 \times 10^{-3} \Msun$/yr).

The inset frames show the density and gas velocity magnitude in a slice as before. The density slice shows a swept-up shell of material that is being radially driven by the star. It also shows the highly asymmetric accretion disk. The radiatively-driven wind shows up in the angles of $20^{\circ}$--$70^{\circ}$, which is the region above the protostellar disk, and $110^{\circ}$--$160^{\circ}$, the region below the disk. Some of these outflows reach 12 km/s.

What is also interesting about this panel is that the inward gas motion at all polar angles has accelerated. This is evidenced by the horizontal line at $v_r = - 6$ km/s. This is gas outside the radiatively-driven bubble undergoing gravitational infall. Within the plane of the disk, $\phi = 90^{\circ} \pm 10^{\circ}$, an enormous amount of gas is seen moving radially inward, mostly with relatively low velocities between $0$ km/s and $-6$ km/s, but some with radial velocities approaching $-20$ km/s. This is accretion flow in the disk. There is also a gas component that has a positive radial velocity within this spread of polar angles with some gas even reaching $8$--$10$ km/s. We expect that the star's own motion relative some component of the gas would produce a positive radial velocity. As the fairly massive disk becomes gravitationally unstable, it also becomes asymmetric. Given that at this stage the disk and stellar masses are still comparable (see Figure \ref{fig:disk_vs_star_mass_evolution}), gravitational interactions between star and disk will displace the star from the center. However, we measure the sink particle speed as only about 1 km/s in the frame of the simulation volume. More likely, some gas orbiting the star in an eccentric elliptical orbit reaches a high radial velocity.

\citet{Seifried+2015} ran simulations of comparable mass (100 $\Msun$) protostellar cores without radiation feedback and found that in both cases, provided that there at least some initial turbulence present, accretion onto the star proceeded along a relatively small number of filamentary channels. Figure \ref{fig:streamlines+toomre} suggests that spiral arms arising from disk instability may aid in focusing accretion along them. This occurs despite not having any initial turbulence in our simulation. Figure \ref{fig:m100_accretion_profile} also suggests that most accretion happens in the plane of the disk.

\section{Discussion} \label{sec:discussion}

The purpose of this paper is to use our powerful new hybrid radiative feedback technique introduced in \citet{Klassen+2014} to understand massive star formation: (1) how do they accrete so much material, (2) what is the role of the disk, (3) does their radiation feedback shut down accretion, and (4) what is the nature of the radiatively-driven bubbles formed by massive stars once they become super-Eddington. All of these questions depend on having a highly accurate radiative feedback method.

Our hybrid radiative transfer scheme is based on a similar one that \citet{Kuiper+2010a} implemented in spherical geometry, also for the study of accretion onto massive stars and their radiative feedback. Our main advance over this scheme is its implementation in a Cartesian geometry with adaptive mesh refinement and its generalization to multiple source terms (stars). We believe that the adoption of hybrid radiative transfer schemes will become more common and it has already been implemented in at least one other AMR code \citep{RamseyDullemond2015}.

After implementing this method in \FLASH, we revisited a problem central to the study of massive stars, but where previous treatments had either left out a treatment of the direct radiation field or were performed in a constrained geometry. The problem focussed on a scaled-up protostellar core with a power-law density profile in slow rigid body rotation. While idealized, it has all the necessary parts to show the role that radiation plays and the relative importance of the disk. 

Despite the absence of any initial turbulence, gravitational instabilities within the protostellar accretion disk eventually destroy the symmetry of the gas distribution. Our analysis of the local Toomre Q parameter for the disk shows how gravitational instability is inevitable as the disk becomes more massive. In each of our three main simulations, the disk went gravitationally unstable in the same way, leading to the formation of spiral arms and an increase in the accretion rate by a factor of 2--10.

Had the simulations included initial turbulence, fragmentation might have present. Observations of giant molecular clouds show supersonic turbulence down to length scales $l > \lambda_s = 0.05$ pc \citep{Ballesteros-Paredes+2007}, where $\lambda_s$ is the sonic scale below which gas motions are subsonic. While supersonic turbulence is essential for providing the density enhancements that result in clumps and cores within GMCs, below the sonic scale subsonic turbulence is less important than thermal pressure in resisting self-gravity and does not contribute to further fragmentation \citep{Padoan1995,Vazquez-Semadeni+2003}. High-mass protostellar objects show line widths in NH$_3$ of about 2.0 km/s \citep{Sridharan+2002}, and molecular cores associated with ultracompact \HII regions show linewidths of 3.1 km/s \citep{Churchwell+1990}. Turbulence is thus associated with high-mass protostellar cores.

Turbulence will likely have an effect on the morphology of radiatively-driven outflow bubble. In our simulations, these bubbles showed a smooth, round morphology, likely because they were expanding into a medium without any local density enhancements. Had the medium been turbulent, the expanding bubble would have engulfed the filaments. Future simulations will explore the relationship to turbulence and radiatively-driven bubbles.

The formation of massive stars is an interplay between gravity and radiation, though with MHD and turbulence playing comparably important roles. For practical reasons, the gray atmosphere (frequency-averaged) approximation is still often used in numerical simulations of radiative feedback, although this is slowly changing as codes become more sophisticated and supercomputers faster. Presently our code uses the gray approximation for both the direct radiation transfer (ray-tracer) and diffusion (FLD). Compared to frequency-dependent approaches, gray radiative transfer will underestimate the optical depth of UV radiation and overestimate the optical depth of infrared radiation. UV radiation will be absorbed closer to the star and infrared radiation will penetrate deeper into the disk, resulting in warmer midplane temperatures \citep{KuiperKlessen2013}.

With magnetic fields, hydromagnetic outflows occur very early \citep[see][simulations of a hydromagnetic disk wind]{BanerjeePudritz2007}. This occurs long before a massive protostellar core even appears. Thus, in the MHD case, the first outflow channel is not created by a radiatively-driven bubble, but by the MHD outflow. This significantly affects bubble evolution in the early phases, and providing an even lower impediment to outflow as studied in \citet{Kuiper+2015} using a sub-grid module for early protostellar outflows. At the same time, magnetic fields are known to suppress disk formation and keep them sub-Keplerian \citep{Seifried+2011, Seifried+2012a}.

Our simulations do not show the formation of any Rayleigh-Taylor type instabilities in the wall of the radiatively-driven bubble. The bubble continues expanding with the shell sweeping up material and deflecting the gravitational infall of new material. Figure \ref{fig:bubble_evolution} shows that material flows along the bubble walls and thus find its way into the accretion disk. Simulations of massive star formation including frequency-dependent irradiation feedback \citep{Kuiper+2010b,Kuiper+2011,Kuiper+2012} show that the mass on top of the cavity wall can already be accelerated into an outflow by the long-wavelength radiation of the stellar spectrum. As the bubble expands, its shell is pushed into regions of lower resolution. We therefore rule out radiative Rayleigh-Taylor instability contributing to disk accretion or direct accretion onto the star.

As a caveat on our simulation, our grid refinement was set using a Jeans length criterion, which resulted in high-density regions being highly refined, and saving the computational cost elsewhere. This, however, meant that as radiatively-driven bubbles expanded, they were not always resolved at the highest level. It is conceivable that the lower resolution suppressed the formation of Rayleigh-Taylor instabilities in the bubble wall. \citet{Krumholz+2009} also had no refinement criteria that focussed explicitly on bubble walls, although they do refine any grid cell where the gradient in the radiation energy density exceeds a 15\% relative value. We are investigating approaches to selectively enhancing the resolution of the bubble wall. Nevertheless, given the fact that the net gas acceleration at the bubble wall is often outward or null, and that momentum appears to be carrying on the expansion even when acceleration is null, we conclude that radiative Rayleigh-Taylor instabilities do not form in the types of environments simulated in this paper.

While radiation pressure may play the dominant role in regulating star formation by disrupting GMCs \citep{Murray+2010}, other important feedback mechanisms are also present and have been left out of the simulations presented in this paper. The other main ways that massive stars disrupt their environments is through the formation of \HII regions and jets.

\HII regions are formed around protostars when the flux of UV photons from the star becomes great enough that a bubble of ionized hydrogen forms. These expanding regions of hot ($10^4$ K), ionized gas are formed around massive stars. Their impact has been studied in various numerical simulations \citep{Peters+2010a,Dale+2014,Walch+2015}, which show that they do not halt accretion onto massive stars, but may act together with stellar winds to reduce the final star formation efficiency. We chose to study radiatively-driven outflows in isolation, but have simulated ionizing feedback in the past \citep{Klassen+2012a, Klassen+2012b}. Ionizing radiation has not yet been fully tested in the most recent version of the \FLASH code, but forthcoming simulations will include ionization feedback and we anticipate that the added thermal pressure would accelerate bubble expansion, though leakage would occur in turbulent environments. The extremely high accretion rates would initially confine any \HII region to the star's immediate vicinity. As optically thin voids form inside the radiatively-driven bubbles, the extreme UV radiation from the massive star would ionize this gas. In our simulations, the sloshing of the high mass disk around the star, as well as the high accretion rates mean that the ionizing radiation is still relatively trapped and the circumstellar disk is not at risk of being photoevaporated (and thus limiting the final mass of the star).

Newly formed protostars and their protostellar disks launch powerful jets and outflows that can carry a significant amount of mass and momentum. Observations show that even low-mass stars launch jets \citep{Dunham+2014}. The launching mechanism for these jets is magnetic in nature \citep{BlandfordPayne1982,PudritzNorman1983,Lynden-Bell2003,BanerjeePudritz2006,Pudritz+2007}, but can be difficult to resolve in simulations of GMC evolution. Subgrid feedback models have been developed that show that jets can reduce the average star mass by a factor of $\sim3$ \citep{Federrath+2014}. The removal of stellar material limits the efficiency of star formation and influences the initial mass function, with \HII regions inflated by radiation pressure predominating in clusters of massive protostars \citep{Fall+2010}.

The launching of jets is fundamentally a magnetic phenomenon and magnetic fields alter the star formation process in important ways that we have not captured in this paper. Magnetic fields suppress star formation through added magnetic pressure support and radiation-magnetohydrodynamic simulations show that magnetic braking of protostellar disks increases radial infall and the accretion luminosity \citep{Commercon+2011b}. The addition of even subsonic turbulence, however, greatly reduces the magnetic braking efficiency \citep{Seifried+2015}.

Our results give an interesting twist to the idea that the accretion of massive stars may be limited by the formation of secondary stars in their gravitationally unstable accretion flows, the so-called fragmentation-induced starvation scenario \citep{Peters+2010a,Peters+2010b,Peters+2010c,Peters+2011}. These models naturally explain that massive stars tend to form in clusters \citep{Peters+2010a}, the observed clustering and number statistics of ultracompact H II regions \citep{Peters+2010b} as well as characteristic properties of poorly collimated high-mass outflows \citep{Peters+2014}. This work is based on larger-scale models compared to our present simulations, with a simulation box of several pc size, a 1000 $\Msun$ initial core, and 98 AU grid resolution. Most importantly, the assumed initial condition is optically thin in the infrared, and the highly optically thick accretion flow around high-mass stars is beyond their resolution limit. Therefore, the radiative heating in these simulations could be treated using raytracing, which gave comparable results to a more accurate Monte Carlo computation \citep{Peters+2010c}. \citet{Peters+2011} speculated that the gravitational fragmentation seen in their radiation-magnetohydrodynamical simulations might continue down to smaller scales, based on the observation that massive stars form with accretion rates of the order $10^{-3} \Msun$/yr, which requires non-Keplerian disks. Our present simulations show that this is indeed the case. However, this gravitational instability does apparently not lead to the formation of companion stars on the smallest scales, so that fragmentation-induced starvation only occurs on scales larger than the disk scale. Future simulations run from different initial conditions will shed further light on this issue.

\section{Conclusion} \label{sec:conclusion}

Numerical simulations of star formation is a truly rich field with many outstanding challenges. Attempting to capture all of the relevant physical processes and assess their relative importance is an ongoing process, and here we simulate some of the most important physical mechanisms that come into play to affect massive star formation. 

We have made important strides in improving existing radiation feedback codes and implementing them in a magnetohydrodynamics code with adaptive mesh refinement, building on the work done by many other authors and collaborators. \citet{Klassen+2014} introduced our hybrid radiative feedback method, blending together the accuracy of a raytracer with the efficiency of a flux-limited diffusion method, as in \citet{Kuiper+2010a} but in a more general 3D Cartesian geometry.

In this paper we have made a major new advance in the study of massive star formation by using this new tool. By simulating protostellar cores of different masses ($30 \Msun$, $100 \Msun$, $200 \Msun$), we showed that even stars with masses over $40 \Msun$ may continue to accrete despite their high luminosity. Through radiation pressure they succeed in driving expanding bubbles that sweep up material and possibly channel gas over their shells back onto the disk. Over the time periods and at the resolution limits we tested, these shells do not show any signs of breaking apart or becoming unstable, but this may still change with the introduction of turbulence and greater resolution in future simulations.

The results presented in this paper are in excellent agreement with recent observations of massive, embedded protostars \citep{BeltrandeWit2015,Johnston+2015}, which show the presence of Keplerian accretion disks of similar radius and mass to what we measure in our simulations. In particular, ALMA observations of AFGL 4176 by \citet{Johnston+2015}, which show a 25 $\Msun$ forming O7-type star surrounded by a 12 $\Msun$ Keplerian-like disk, strongly resemble the results of our 100 $\Msun$ simulation.

Protostellar disks grow rapidly and become Toomre unstable. These instabilities do not result in further star formation, but instead form spiral arms and an assymetric disk that channels material onto the star at rates 2--10 times faster than before. The disk mass begins to level out at around this same time while the stellar mass continues to grow.

We now summarize the results of our investigation as follows:
\begin{itemize}
\item Each of our three simulated massive protostellar cores produced only a single star, despite a Toomre analysis showing their circumstellar disks to be unstable.
\item After 81.4 kyr of evolution, our $30 \Msun$ simulation showed a star with a mass of $5.48 \Msun$, while our $100 \Msun$ simulation formed a $28.84 \Msun$ mass star and our $200 \Msun$ simulation formed a $43.65 \Msun$ star. The latter two were about 30 and 100 times super-Eddington in their luminosity, respectively, and drove powerful winds occassionally achieving speeds of greater than 50 km/s.
\item Despite becoming locally Toomre Q unstable and forming spiral arms, the accretion disks do not fragment gravitationally to form more stars, at least for the duration of our simulations. We use the Hill criterion analysis of \citet{RogersWadsley2012} to show that these spiral arms are still stable against fragmentation. We also find that the combined \citet{Gammie2001} and Toomre conditions for fragmentation predict fragmentation of the disk to potentially form a binary companion at $\sim 1200$ AU radius, but this has not (at least, yet) occurred.
\item Accretion onto massive protostars occurs very efficiently through a protostellar disk, despite an extremely high flux of photons. Towards the ends of our simulations, accretion rates were about $5\times10^{-5} \Msun$/yr, $6 \times 10^{-4} \Msun$/yr, and $1.5\times10^{-3} \Msun$/yr for the $30$, $100$, and $200 \Msun$ simulations, respectively. These showed no signs of slowing down significantly and we speculate that our stars could have gone on to accrete a significant fraction of the total core mass.
\item Optically-thick circumstellar disks are responsible for the flashlight effect, i.e. the channeling of flux along the polar axis and into the polar cavities. Dust sublimation reduces the optical depth, but also allows radiation to more easily escape into the polar cavities.
\item After the stars luminosities exceeded the Eddington limit, they drove radiative bubbles, sometimes in successive shells that could appear pierced by stellar winds.
\item Material was observed flowing along the outer shell of these bubbles back onto the circumstellar disk, but the outer shell itself did not show signs of a Rayleigh-Taylor instability. Net gas acceleration at the bubble wall is outward or close to zero. We conclude that radiative Rayleigh-Taylor instabilities do not form in the types of environments simulated in this paper.
\end{itemize}

Future simulations will explore the effects of turbulence, ionizing feedback, and magnetic fields, each of which contributes to the star formation efficiency and cloud lifetime in important ways. Although our model is idealized, it allows us to investigate several key processes in great detail that would have been difficult to tease apart if more physics had been included in our model. The way forward will be to look at each in turn.

\section*{Acknowledgments}

We thank the referee for providing a very useful report that helped improve the paper. The development of this hybrid radiation transfer code benefited from helpful discussions with Romain Teyssier, Christoph Federrath, Klaus Weide, and Petros Tzeferacos, and from two visits by M.K.~to the \FLASH Center at the University of Chicago. We also thank Henrik Beuther, Thomas Henning, and Ralf Klessen for interesting discussions about observations and theory. Manos Chatzopoulos supplied the modified unsplit hydrodynamics solver that improved the accuracy of our simulations. M.K.~acknowledges financial support from the National Sciences and Engineering Research Council (NSERC) of Canada through a Postgraduate Scholarship. R.E.P.~is supported by an NSERC Discovery Grant. R.K.~acknowledges financial support by the Emmy-Noether-Program of the German Research Foundation (DFG) under grant no. KU 2849/3-1. T.P. acknowledges support from the DFG Priority Program 1573 {\em Physics of the Interstellar Medium}. The \FLASH code was in part developed by the DOE-supported Alliances Center for Astrophysical Thermonuclear Flashes (ASCI) at the University of Chicago\footnote{\url{http://flash.uchicago.edu/}}. This work was made possible by the facilities of the Shared Hierarchical Academic Research Computing Network (SHARCNET: www.sharcnet.ca), SciNet at the University of Toronto, and Compute/Calcul Canada.

We are grateful to KITP, Santa Barbara, for supporting M.K.~as an affiliate (for 3 weeks), R.E.P.~as an invited participant (for 2.5 months), and R.K.~as an invited participant (1 week) in the program ``Gravity's Loyal Opposition: The Physics of Star Formation Feedback,'' where some of this research was supported by the National Science Foundation under Grant No. NSF PHY11-25915. R.E.P.~also thanks the MPIA and the Institut f\"{u}r Theoretische Astrophysik (ITA) in the Zentrum f\"{u}r Astronomie Heidelberg for support during his sabbatical leave (2015/16) during the final stages of this project. 

Much of the analysis and data visualization was performed using the {\tt yt} toolkit\footnote{\url{http://yt-project.org}} by \citet{ytpaper}. 

\bibliography{radiation_pressure_cores}

\begin{thebibliography}{107}
\expandafter\ifx\csname natexlab\endcsname\relax\def\natexlab#1{#1}\fi

\bibitem[{Amanatides {et~al.}(1987)Amanatides, Woo,
  {et~al.}}]{AmanatidesWoo1987}
Amanatides, J., Woo, A., {et~al.} 1987, in Proceedings of EUROGRAPHICS,
  Vol.~87, 3--10

\bibitem[{{Ballesteros-Paredes} {et~al.}(2007){Ballesteros-Paredes}, {Klessen},
  {Mac Low}, \& {Vazquez-Semadeni}}]{Ballesteros-Paredes+2007}
{Ballesteros-Paredes}, J., {Klessen}, R.~S., {Mac Low}, M.-M., \&
  {Vazquez-Semadeni}, E. 2007, Protostars and Planets V, 63

\bibitem[{{Banerjee} \& {Pudritz}(2006)}]{BanerjeePudritz2006}
{Banerjee}, R., \& {Pudritz}, R.~E. 2006, \apj, 641, 949

\bibitem[{{Banerjee} \& {Pudritz}(2007)}]{BanerjeePudritz2007}
---. 2007, \apj, 660, 479

\bibitem[{{Beltr{\'a}n} {et~al.}(2006){Beltr{\'a}n}, {Brand}, {Cesaroni},
  {Fontani}, {Pezzuto}, {Testi}, \& {Molinari}}]{Beltran+2006}
{Beltr{\'a}n}, M.~T., {Brand}, J., {Cesaroni}, R., {et~al.} 2006, \aap, 447,
  221

\bibitem[{{Beltran} \& {de Wit}(2015)}]{BeltrandeWit2015}
{Beltran}, M.~T., \& {de Wit}, W.~J. 2015, ArXiv e-prints

\bibitem[{{Blandford} \& {Payne}(1982)}]{BlandfordPayne1982}
{Blandford}, R.~D., \& {Payne}, D.~G. 1982, \mnras, 199, 883

\bibitem[{{Buntemeyer} {et~al.}(2015){Buntemeyer}, {Banerjee}, {Peters},
  {Klassen}, \& {Pudritz}}]{Buntemeyer+2015}
{Buntemeyer}, L., {Banerjee}, R., {Peters}, T., {Klassen}, M., \& {Pudritz},
  R.~E. 2015, ArXiv e-prints

\bibitem[{{Chini} {et~al.}(2012){Chini}, {Hoffmeister}, {Nasseri}, {Stahl}, \&
  {Zinnecker}}]{Chini+2012}
{Chini}, R., {Hoffmeister}, V.~H., {Nasseri}, A., {Stahl}, O., \& {Zinnecker},
  H. 2012, \mnras, 424, 1925

\bibitem[{{Churchwell} {et~al.}(1990){Churchwell}, {Walmsley}, \&
  {Cesaroni}}]{Churchwell+1990}
{Churchwell}, E., {Walmsley}, C.~M., \& {Cesaroni}, R. 1990, \aaps, 83, 119

\bibitem[{{Colella} \& {Woodward}(1984)}]{ColellaWoodward1984}
{Colella}, P., \& {Woodward}, P.~R. 1984, Journal of Computational Physics, 54,
  174

\bibitem[{{Commer{\c c}on} {et~al.}(2011{\natexlab{a}}){Commer{\c c}on},
  {Hennebelle}, \& {Henning}}]{Commercon+2011b}
{Commer{\c c}on}, B., {Hennebelle}, P., \& {Henning}, T. 2011{\natexlab{a}},
  \apjl, 742, L9

\bibitem[{{Commer{\c c}on} {et~al.}(2011{\natexlab{b}}){Commer{\c c}on},
  {Teyssier}, {Audit}, {Hennebelle}, \& {Chabrier}}]{Commercon+2011}
{Commer{\c c}on}, B., {Teyssier}, R., {Audit}, E., {Hennebelle}, P., \&
  {Chabrier}, G. 2011{\natexlab{b}}, \aap, 529, A35

\bibitem[{{Crowther} {et~al.}(2010){Crowther}, {Schnurr}, {Hirschi}, {Yusof},
  {Parker}, {Goodwin}, \& {Kassim}}]{Crowther+2010}
{Crowther}, P.~A., {Schnurr}, O., {Hirschi}, R., {et~al.} 2010, \mnras, 408,
  731

\bibitem[{{Dale} {et~al.}(2014){Dale}, {Ngoumou}, {Ercolano}, \&
  {Bonnell}}]{Dale+2014}
{Dale}, J.~E., {Ngoumou}, J., {Ercolano}, B., \& {Bonnell}, I.~A. 2014, \mnras,
  442, 694

\bibitem[{{Doran} {et~al.}(2013){Doran}, {Crowther}, {de Koter}, {Evans},
  {McEvoy}, {Walborn}, {Bastian}, {Bestenlehner}, {Gr{\"a}fener}, {Herrero},
  {K{\"o}hler}, {Ma{\'{\i}}z Apell{\'a}niz}, {Najarro}, {Puls}, {Sana},
  {Schneider}, {Taylor}, {van Loon}, \& {Vink}}]{Doran+2013}
{Doran}, E.~I., {Crowther}, P.~A., {de Koter}, A., {et~al.} 2013, \aap, 558,
  A134

\bibitem[{{Draine} \& {Lee}(1984)}]{DraineLee1984}
{Draine}, B.~T., \& {Lee}, H.~M. 1984, \apj, 285, 89

\bibitem[{Dubey {et~al.}(2009)Dubey, Antypas, Ganapathy, Reid, Riley, Sheeler,
  Siegel, \& Weide}]{Dubey+2009}
Dubey, A., Antypas, K., Ganapathy, M.~K., {et~al.} 2009, Parallel Computing,
  35, 512

\bibitem[{{Dunham} {et~al.}(2014){Dunham}, {Arce}, {Mardones}, {Lee},
  {Matthews}, {Stutz}, \& {Williams}}]{Dunham+2014}
{Dunham}, M.~M., {Arce}, H.~G., {Mardones}, D., {et~al.} 2014, \apj, 783, 29

\bibitem[{{Edgar} \& {Clarke}(2003)}]{EdgarClarke2003}
{Edgar}, R., \& {Clarke}, C. 2003, \mnras, 338, 962

\bibitem[{{Evans}(1999)}]{Evans1999}
{Evans}, II, N.~J. 1999, \araa, 37, 311

\bibitem[{Falgout \& Yang(2002)}]{HYPRE}
Falgout, R., \& Yang, U. 2002, in Lecture Notes in Computer Science, Vol. 2331,
  Computational Science --€" ICCS 2002, ed. P.~Sloot, A.~Hoekstra, C.~Tan, \&
  J.~Dongarra (Springer Berlin Heidelberg), 632--641

\bibitem[{{Fall} {et~al.}(2010){Fall}, {Krumholz}, \& {Matzner}}]{Fall+2010}
{Fall}, S.~M., {Krumholz}, M.~R., \& {Matzner}, C.~D. 2010, \apjl, 710, L142

\bibitem[{{Federrath} {et~al.}(2010){Federrath}, {Banerjee}, {Clark}, \&
  {Klessen}}]{Federrath+2010}
{Federrath}, C., {Banerjee}, R., {Clark}, P.~C., \& {Klessen}, R.~S. 2010,
  \apj, 713, 269

\bibitem[{{Federrath} \& {Klessen}(2012)}]{FederrathKlessen2012}
{Federrath}, C., \& {Klessen}, R.~S. 2012, \apj, 761, 156

\bibitem[{{Federrath} {et~al.}(2014){Federrath}, {Schr{\"o}n}, {Banerjee}, \&
  {Klessen}}]{Federrath+2014}
{Federrath}, C., {Schr{\"o}n}, M., {Banerjee}, R., \& {Klessen}, R.~S. 2014,
  \apj, 790, 128

\bibitem[{{Figer}(2005)}]{Figer2005}
{Figer}, D.~F. 2005, \nat, 434, 192

\bibitem[{{Fryxell} {et~al.}(2000){Fryxell}, {Olson}, {Ricker}, {Timmes},
  {Zingale}, {Lamb}, {MacNeice}, {Rosner}, {Truran}, \& {Tufo}}]{Fryxell+2000}
{Fryxell}, B., {Olson}, K., {Ricker}, P., {et~al.} 2000, \apjs, 131, 273

\bibitem[{{Gammie}(2001)}]{Gammie2001}
{Gammie}, C.~F. 2001, \apj, 553, 174

\bibitem[{{Girichidis} {et~al.}(2011){Girichidis}, {Federrath}, {Banerjee}, \&
  {Klessen}}]{Girichidis+2011}
{Girichidis}, P., {Federrath}, C., {Banerjee}, R., \& {Klessen}, R.~S. 2011,
  \mnras, 413, 2741

\bibitem[{{Goodwin} \& {Kroupa}(2005)}]{GoodwinKroupa2005}
{Goodwin}, S.~P., \& {Kroupa}, P. 2005, \aap, 439, 565

\bibitem[{{Goodwin} {et~al.}(2004){Goodwin}, {Whitworth}, \&
  {Ward-Thompson}}]{Goodwin+2004}
{Goodwin}, S.~P., {Whitworth}, A.~P., \& {Ward-Thompson}, D. 2004, \aap, 414,
  633

\bibitem[{{Haemmerl{\'e}} \& {Peters}(2016)}]{HaemmerlePeters2016}
{Haemmerl{\'e}}, L., \& {Peters}, T. 2016, \mnras

\bibitem[{{Harries}(2015)}]{Harries2015}
{Harries}, T.~J. 2015, \mnras, 448, 3156

\bibitem[{{Harries} {et~al.}(2014){Harries}, {Haworth}, \&
  {Acreman}}]{Harries+2014}
{Harries}, T.~J., {Haworth}, T.~J., \& {Acreman}, D.~M. 2014, Astrophysics and
  Space Science Proceedings, 36, 395

\bibitem[{{Hill} {et~al.}(2005){Hill}, {Burton}, {Minier}, {Thompson}, {Walsh},
  {Hunt-Cunningham}, \& {Garay}}]{Hill+2005}
{Hill}, T., {Burton}, M.~G., {Minier}, V., {et~al.} 2005, \mnras, 363, 405

\bibitem[{{Hosokawa} \& {Omukai}(2009)}]{HosokawaOmukai2009}
{Hosokawa}, T., \& {Omukai}, K. 2009, \apj, 691, 823

\bibitem[{{Howard} {et~al.}(2014){Howard}, {Pudritz}, \&
  {Harris}}]{Howard+2014}
{Howard}, C.~S., {Pudritz}, R.~E., \& {Harris}, W.~E. 2014, \mnras, 438, 1305

\bibitem[{{Isella} \& {Natta}(2005)}]{IsellaNatta2005}
{Isella}, A., \& {Natta}, A. 2005, \aap, 438, 899

\bibitem[{{Jacquet} \& {Krumholz}(2011)}]{JacquetKrumholz2011}
{Jacquet}, E., \& {Krumholz}, M.~R. 2011, \apj, 730, 116

\bibitem[{{Johnson} \& {Gammie}(2003)}]{JohnsonGammie2003}
{Johnson}, B.~M., \& {Gammie}, C.~F. 2003, \apj, 597, 131

\bibitem[{{Johnston} {et~al.}(2015){Johnston}, {Robitaille}, {Beuther}, {Linz},
  {Boley}, {Kuiper}, {Keto}, {Hoare}, \& {van Boekel}}]{Johnston+2015}
{Johnston}, K.~G., {Robitaille}, T.~P., {Beuther}, H., {et~al.} 2015, ArXiv
  e-prints

\bibitem[{{Kahn}(1974)}]{Kahn1974}
{Kahn}, F.~D. 1974, \aap, 37, 149

\bibitem[{{Klassen} {et~al.}(2014){Klassen}, {Kuiper}, {Pudritz}, {Peters},
  {Banerjee}, \& {Buntemeyer}}]{Klassen+2014}
{Klassen}, M., {Kuiper}, R., {Pudritz}, R.~E., {et~al.} 2014, \apj, 797, 4

\bibitem[{{Klassen} {et~al.}(2012{\natexlab{a}}){Klassen}, {Peters}, \&
  {Pudritz}}]{Klassen+2012b}
{Klassen}, M., {Peters}, T., \& {Pudritz}, R.~E. 2012{\natexlab{a}}, \apj, 758,
  137

\bibitem[{{Klassen} {et~al.}(2012{\natexlab{b}}){Klassen}, {Pudritz}, \&
  {Peters}}]{Klassen+2012a}
{Klassen}, M., {Pudritz}, R.~E., \& {Peters}, T. 2012{\natexlab{b}}, \mnras,
  421, 2861

\bibitem[{{Klein} {et~al.}(2005){Klein}, {Posselt}, {Schreyer}, {Forbrich}, \&
  {Henning}}]{Klein+2005}
{Klein}, R., {Posselt}, B., {Schreyer}, K., {Forbrich}, J., \& {Henning}, T.
  2005, \apjs, 161, 361

\bibitem[{{Krumholz} {et~al.}(2007){Krumholz}, {Klein}, {McKee}, \&
  {Bolstad}}]{Krumholz+2007b}
{Krumholz}, M.~R., {Klein}, R.~I., {McKee}, C.~F., \& {Bolstad}, J. 2007, \apj,
  667, 626

\bibitem[{{Krumholz} {et~al.}(2009){Krumholz}, {Klein}, {McKee}, {Offner}, \&
  {Cunningham}}]{Krumholz+2009}
{Krumholz}, M.~R., {Klein}, R.~I., {McKee}, C.~F., {Offner}, S.~S.~R., \&
  {Cunningham}, A.~J. 2009, Science, 323, 754

\bibitem[{{Krumholz} {et~al.}(2004){Krumholz}, {McKee}, \&
  {Klein}}]{Krumholz+2004}
{Krumholz}, M.~R., {McKee}, C.~F., \& {Klein}, R.~I. 2004, \apj, 611, 399

\bibitem[{{Kuiper} {et~al.}(2010{\natexlab{a}}){Kuiper}, {Klahr}, {Beuther}, \&
  {Henning}}]{Kuiper+2010b}
{Kuiper}, R., {Klahr}, H., {Beuther}, H., \& {Henning}, T. 2010{\natexlab{a}},
  \apj, 722, 1556

\bibitem[{{Kuiper} {et~al.}(2011){Kuiper}, {Klahr}, {Beuther}, \&
  {Henning}}]{Kuiper+2011}
---. 2011, \apj, 732, 20

\bibitem[{{Kuiper} {et~al.}(2012){Kuiper}, {Klahr}, {Beuther}, \&
  {Henning}}]{Kuiper+2012}
---. 2012, \aap, 537, A122

\bibitem[{{Kuiper} {et~al.}(2010{\natexlab{b}}){Kuiper}, {Klahr}, {Dullemond},
  {Kley}, \& {Henning}}]{Kuiper+2010a}
{Kuiper}, R., {Klahr}, H., {Dullemond}, C., {Kley}, W., \& {Henning}, T.
  2010{\natexlab{b}}, \aap, 511, A81

\bibitem[{{Kuiper} \& {Klessen}(2013)}]{KuiperKlessen2013}
{Kuiper}, R., \& {Klessen}, R.~S. 2013, \aap, 555, A7

\bibitem[{{Kuiper} \& {Yorke}(2013{\natexlab{a}})}]{KuiperYorke2013a}
{Kuiper}, R., \& {Yorke}, H.~W. 2013{\natexlab{a}}, \apj, 763, 104

\bibitem[{{Kuiper} \& {Yorke}(2013{\natexlab{b}})}]{KuiperYorke2013b}
---. 2013{\natexlab{b}}, \apj, 772, 61

\bibitem[{{Kuiper} {et~al.}(2015){Kuiper}, {Yorke}, \& {Turner}}]{Kuiper+2015}
{Kuiper}, R., {Yorke}, H.~W., \& {Turner}, N.~J. 2015, \apj, 800, 86

\bibitem[{{Larson}(1981)}]{Larson1981}
{Larson}, R.~B. 1981, \mnras, 194, 809

\bibitem[{{Larson} \& {Starrfield}(1971)}]{LarsonStarrfield1971}
{Larson}, R.~B., \& {Starrfield}, S. 1971, \aap, 13, 190

\bibitem[{{Levermore} \& {Pomraning}(1981)}]{LevermorePomraning1981}
{Levermore}, C.~D., \& {Pomraning}, G.~C. 1981, \apj, 248, 321

\bibitem[{{Lynden-Bell}(2003)}]{Lynden-Bell2003}
{Lynden-Bell}, D. 2003, \mnras, 341, 1360

\bibitem[{{Mac Low} \& {Klessen}(2004)}]{MacLowKlessen2004}
{Mac Low}, M.-M., \& {Klessen}, R.~S. 2004, Reviews of Modern Physics, 76, 125

\bibitem[{{MacNeice} {et~al.}(2000){MacNeice}, {Olson}, {Mobarry}, {de
  Fainchtein}, \& {Packer}}]{MacNeice+2000}
{MacNeice}, P., {Olson}, K.~M., {Mobarry}, C., {de Fainchtein}, R., \&
  {Packer}, C. 2000, Computer Physics Communications, 126, 330

\bibitem[{{McKee} \& {Ostriker}(2007)}]{McKeeOstriker2007}
{McKee}, C.~F., \& {Ostriker}, E.~C. 2007, \araa, 45, 565

\bibitem[{{Mihalas} \& {Mihalas}(1984)}]{MihalasMihalas84}
{Mihalas}, D., \& {Mihalas}, B.~W. 1984, {Foundations of radiation
  hydrodynamics} (Oxford University Press)

\bibitem[{{Minerbo}(1978)}]{Minerbo1978}
{Minerbo}, G.~N. 1978, \jqsrt, 20, 541

\bibitem[{{Murray} {et~al.}(2011){Murray}, {M{\'e}nard}, \&
  {Thompson}}]{Murray+2011}
{Murray}, N., {M{\'e}nard}, B., \& {Thompson}, T.~A. 2011, \apj, 735, 66

\bibitem[{{Murray} {et~al.}(2010){Murray}, {Quataert}, \&
  {Thompson}}]{Murray+2010}
{Murray}, N., {Quataert}, E., \& {Thompson}, T.~A. 2010, \apj, 709, 191

\bibitem[{{Murray} {et~al.}(1994){Murray}, {Castor}, {Klein}, \&
  {McKee}}]{Murray+1994}
{Murray}, S.~D., {Castor}, J.~I., {Klein}, R.~I., \& {McKee}, C.~F. 1994, \apj,
  435, 631

\bibitem[{{Nakano}(1989)}]{Nakano1989}
{Nakano}, T. 1989, \apj, 345, 464

\bibitem[{{Ochsendorf} {et~al.}(2014){Ochsendorf}, {Verdolini}, {Cox},
  {Bern{\'e}}, {Kaper}, \& {Tielens}}]{Ochsendorf+2014}
{Ochsendorf}, B.~B., {Verdolini}, S., {Cox}, N.~L.~J., {et~al.} 2014, \aap,
  566, A75

\bibitem[{{Offner} {et~al.}(2009){Offner}, {Klein}, {McKee}, \&
  {Krumholz}}]{Offner+2009}
{Offner}, S.~S.~R., {Klein}, R.~I., {McKee}, C.~F., \& {Krumholz}, M.~R. 2009,
  \apj, 703, 131

\bibitem[{{Padoan}(1995)}]{Padoan1995}
{Padoan}, P. 1995, \mnras, 277, 377

\bibitem[{{Peters} {et~al.}(2011){Peters}, {Banerjee}, {Klessen}, \& {Mac
  Low}}]{Peters+2011}
{Peters}, T., {Banerjee}, R., {Klessen}, R.~S., \& {Mac Low}, M.-M. 2011, \apj,
  729, 72

\bibitem[{{Peters} {et~al.}(2010{\natexlab{a}}){Peters}, {Banerjee}, {Klessen},
  {Mac Low}, {Galv{\'a}n-Madrid}, \& {Keto}}]{Peters+2010a}
{Peters}, T., {Banerjee}, R., {Klessen}, R.~S., {et~al.} 2010{\natexlab{a}},
  \apj, 711, 1017

\bibitem[{{Peters} {et~al.}(2014){Peters}, {Klaassen}, {Mac Low}, {Schr{\"o}n},
  {Federrath}, {Smith}, \& {Klessen}}]{Peters+2014}
{Peters}, T., {Klaassen}, P.~D., {Mac Low}, M.-M., {et~al.} 2014, \apj, 788, 14

\bibitem[{{Peters} {et~al.}(2010{\natexlab{b}}){Peters}, {Klessen}, {Mac Low},
  \& {Banerjee}}]{Peters+2010c}
{Peters}, T., {Klessen}, R.~S., {Mac Low}, M.-M., \& {Banerjee}, R.
  2010{\natexlab{b}}, \apj, 725, 134

\bibitem[{{Peters} {et~al.}(2010{\natexlab{c}}){Peters}, {Mac Low}, {Banerjee},
  {Klessen}, \& {Dullemond}}]{Peters+2010b}
{Peters}, T., {Mac Low}, M.-M., {Banerjee}, R., {Klessen}, R.~S., \&
  {Dullemond}, C.~P. 2010{\natexlab{c}}, \apj, 719, 831

\bibitem[{{Pollack} {et~al.}(1994){Pollack}, {Hollenbach}, {Beckwith},
  {Simonelli}, {Roush}, \& {Fong}}]{Pollack+1994}
{Pollack}, J.~B., {Hollenbach}, D., {Beckwith}, S., {et~al.} 1994, \apj, 421,
  615

\bibitem[{{Pringle}(1996)}]{Pringle1996}
{Pringle}, J.~E. 1996, \mnras, 281, 357

\bibitem[{{Pudritz} \& {Norman}(1983)}]{PudritzNorman1983}
{Pudritz}, R.~E., \& {Norman}, C.~A. 1983, \apj, 274, 677

\bibitem[{{Pudritz} {et~al.}(2007){Pudritz}, {Ouyed}, {Fendt}, \&
  {Brandenburg}}]{Pudritz+2007}
{Pudritz}, R.~E., {Ouyed}, R., {Fendt}, C., \& {Brandenburg}, A. 2007,
  Protostars and Planets V, 277

\bibitem[{{Ramsey} \& {Dullemond}(2015)}]{RamseyDullemond2015}
{Ramsey}, J.~P., \& {Dullemond}, C.~P. 2015, \aap, 574, A81

\bibitem[{{Rogers} \& {Wadsley}(2012)}]{RogersWadsley2012}
{Rogers}, P.~D., \& {Wadsley}, J. 2012, \mnras, 423, 1896

\bibitem[{Saad \& Schultz(1986)}]{GMRES}
Saad, Y., \& Schultz, M.~H. 1986, SIAM J. Sci. Stat. Comput., 7, 856

\bibitem[{{Seifried} {et~al.}(2011){Seifried}, {Banerjee}, {Klessen}, {Duffin},
  \& {Pudritz}}]{Seifried+2011}
{Seifried}, D., {Banerjee}, R., {Klessen}, R.~S., {Duffin}, D., \& {Pudritz},
  R.~E. 2011, \mnras, 417, 1054

\bibitem[{{Seifried} {et~al.}(2015){Seifried}, {Banerjee}, {Pudritz}, \&
  {Klessen}}]{Seifried+2015}
{Seifried}, D., {Banerjee}, R., {Pudritz}, R.~E., \& {Klessen}, R.~S. 2015,
  \mnras, 446, 2776

\bibitem[{{Seifried} {et~al.}(2012){Seifried}, {Pudritz}, {Banerjee}, {Duffin},
  \& {Klessen}}]{Seifried+2012a}
{Seifried}, D., {Pudritz}, R.~E., {Banerjee}, R., {Duffin}, D., \& {Klessen},
  R.~S. 2012, \mnras, 422, 347

\bibitem[{{Shu}(1977)}]{Shu1977}
{Shu}, F.~H. 1977, \apj, 214, 488

\bibitem[{{Sridharan} {et~al.}(2002){Sridharan}, {Beuther}, {Schilke},
  {Menten}, \& {Wyrowski}}]{Sridharan+2002}
{Sridharan}, T.~K., {Beuther}, H., {Schilke}, P., {Menten}, K.~M., \&
  {Wyrowski}, F. 2002, \apj, 566, 931

\bibitem[{{Stahler} {et~al.}(2000){Stahler}, {Palla}, \& {Ho}}]{Stahler+2000}
{Stahler}, S.~W., {Palla}, F., \& {Ho}, P.~T.~P. 2000, Protostars and Planets
  IV, 327

\bibitem[{{Takahashi} {et~al.}(2016){Takahashi}, {Tsukamoto}, \&
  {Inutsuka}}]{Takahashi+2016}
{Takahashi}, S.~Z., {Tsukamoto}, Y., \& {Inutsuka}, S.-i. 2016, ArXiv e-prints

\bibitem[{{Tenorio-Tagle}(1979)}]{Tenorio-Tagle1979}
{Tenorio-Tagle}, G. 1979, \aap, 71, 59

\bibitem[{{Toomre}(1964)}]{Toomre1964}
{Toomre}, A. 1964, \apj, 139, 1217

\bibitem[{{Truelove} {et~al.}(1997){Truelove}, {Klein}, {McKee}, {Holliman},
  {Howell}, \& {Greenough}}]{Truelove+1997}
{Truelove}, J.~K., {Klein}, R.~I., {McKee}, C.~F., {et~al.} 1997, \apjl, 489,
  L179

\bibitem[{{Trujillo Bueno} \& {Fabiani Bendicho}(1995)}]{ALI1995}
{Trujillo Bueno}, J., \& {Fabiani Bendicho}, P. 1995, \apj, 455, 646

\bibitem[{{Turk} {et~al.}(2011){Turk}, {Smith}, {Oishi}, {Skory}, {Skillman},
  {Abel}, \& {Norman}}]{ytpaper}
{Turk}, M.~J., {Smith}, B.~D., {Oishi}, J.~S., {et~al.} 2011, \apjs, 192, 9

\bibitem[{{Turner} \& {Stone}(2001)}]{TurnerStone2001}
{Turner}, N.~J., \& {Stone}, J.~M. 2001, \apjs, 135, 95

\bibitem[{{V{\'a}zquez-Semadeni} {et~al.}(2003){V{\'a}zquez-Semadeni},
  {Ballesteros-Paredes}, \& {Klessen}}]{Vazquez-Semadeni+2003}
{V{\'a}zquez-Semadeni}, E., {Ballesteros-Paredes}, J., \& {Klessen}, R.~S.
  2003, \apjl, 585, L131

\bibitem[{{Walch} {et~al.}(2015){Walch}, {Whitworth}, {Bisbas}, {Hubber}, \&
  {W{\"u}nsch}}]{Walch+2015}
{Walch}, S., {Whitworth}, A.~P., {Bisbas}, T.~G., {Hubber}, D.~A., \&
  {W{\"u}nsch}, R. 2015, \mnras, 452, 2794

\bibitem[{{Wolfire} \& {Cassinelli}(1986)}]{WolfireCassinelli1986}
{Wolfire}, M.~G., \& {Cassinelli}, J.~P. 1986, \apj, 310, 207

\bibitem[{{Yorke} \& {Bodenheimer}(1999)}]{YorkeBodenheimer1999}
{Yorke}, H.~W., \& {Bodenheimer}, P. 1999, \apj, 525, 330

\bibitem[{{Yorke} \& {Kr\"{u}gel}(1977)}]{YorkeKruegel1977}
{Yorke}, H.~W., \& {Kr\"{u}gel}, E. 1977, \aap, 54, 183

\bibitem[{{Yorke} \& {Sonnhalter}(2002)}]{YorkeSonnhalter2002}
{Yorke}, H.~W., \& {Sonnhalter}, C. 2002, \apj, 569, 846

\bibitem[{{Zhang} {et~al.}(2011){Zhang}, {Howell}, {Almgren}, {Burrows}, \&
  {Bell}}]{Zhang+2011}
{Zhang}, W., {Howell}, L., {Almgren}, A., {Burrows}, A., \& {Bell}, J. 2011,
  \apjs, 196, 20

\bibitem[{{Zinnecker} \& {Yorke}(2007)}]{ZinneckerYorke2007}
{Zinnecker}, H., \& {Yorke}, H.~W. 2007, \araa, 45, 481

\end{thebibliography}

\label{lastpage}

\end{document}